\documentclass{emulateapj}
\usepackage{verbatim,graphicx,natbib}
\shortauthors{Cowan, Voigt \& Abbot}
\shorttitle{Thermal Phases of Earth-Like Planets}
\begin{document}

\title{Thermal Phases of Earth-Like Planets: Estimating Thermal Inertia from Eccentricity, Obliquity, and Diurnal Forcing}
\author{Nicolas B. Cowan\altaffilmark{1,2}, Aiko Voigt\altaffilmark{3}, Dorian S. Abbot\altaffilmark{4}}
\altaffiltext{1}{Center for Interdisciplinary Exploration and Research in Astrophysics and Department of Physics \& Astronomy, Northwestern University, 2131 Tech Drive, Evanston, IL 60208, USA}
\altaffiltext{2}{CIERA Postdoctoral Fellow}
\altaffiltext{3}{Max Planck Institute for Meteorology, Bundesstr. 53, D-20146, Hamburg, Germany}
\altaffiltext{4}{Department of Geophysical Sciences, University of Chicago, 5734 South Ellis Avenue, Chicago, IL, 60637, USA}

\begin{abstract}
In order to understand the climate on terrestrial planets orbiting nearby Sun-like stars, one would like to know their thermal inertia. We use a global climate model to simulate the thermal phase variations of Earth-analogs and test whether these data could distinguish between planets with different heat storage and heat transport characteristics. In particular, we consider a temperate climate with polar ice caps (like modern Earth), and a snowball state where the oceans are globally covered in ice.  We first quantitatively study the periodic radiative forcing from, and climatic response to, rotation, obliquity, and eccentricity.  Orbital eccentricity and seasonal changes in albedo cause variations in the global-mean absorbed flux. The responses of the two climates to these global seasons indicate that the temperate planet has 3$\times$ the bulk heat capacity of the snowball planet due to the presence of liquid water oceans.  The temperate obliquity seasons are weaker than one would expect based on thermal inertia alone; this is due to cross-equatorial oceanic and atmospheric energy transport. Thermal inertia and cross-equatorial heat transport have qualitatively different effects on obliquity seasons, insofar as heat transport tends to reduce seasonal amplitude without inducing a phase lag.  For an Earth-like planet, however, this effect is masked by the mixing of signals from low thermal inertia regions (sea ice and land) with that from high thermal inertia regions (oceans), which also produces a damped response with small phase lag.  We then simulate thermal lightcurves as they would appear to a high-contrast imaging mission (TPF-I/Darwin).
In order of importance to the present simulations, which use modern-Earth orbital parameters, the three drivers of thermal phase variations are 1) obliquity seasons, 2) diurnal cycle, and 3) global seasons. Obliquity seasons are the dominant source of phase variations for most viewing angles. A pole-on observer would measure peak-to-trough amplitudes of 13\% and 47\% for the temperate and snowball climates, respectively. Diurnal heating is important for equatorial observers ($\sim5$\% phase variations), because the obliquity effects cancel to first order from that vantage.  Finally, we compare the prospects of optical vs.\ thermal direct imaging missions for constraining the climate on exoplanets and conclude that while zero- and one-dimensional models are best served by thermal measurements, second-order models accounting for seasons and planetary thermal inertia would require \emph{both} optical and thermal observations.
\end{abstract}

\section{Introduction}
An extrasolar planet's disk-averaged emitting temperature changes throughout its orbit, a phenomenon known as thermal phase variations.  In this paper we use a comprehensive global climate model (GCM) to compare the thermal phase variations for two very different climates that are known to have existed in the Sun's habitable zone (HZ): a temperate climate, and a pan-glacial climate from Earth's geological past (``snowball Earth'').  After introducing our climate model (Section ~2) and discussing its time-averaged characteristics (Section~3), we study the periodic radiative forcings and intrinsic climate response (Section~4). We then consider, in Section~5, how these climate variations manifest themselves to an external observer; we discuss our results in Section~6.  Such a study can help determine whether time-resolved thermal photometry will be as useful a tool for directly imaged exoplanets as it is for the current crop of short-period giant planets.

As \cite{Cowan_2011c} noted, the most direct way to distinguish a temperate climate from a snowball is to measure the planet's Bond albedo. By combining an exoplanet's emitting temperature with its bolometric flux, one can obtain estimates of its radius and albedo.  The albedo of a snowball planet would be more than twice that of a terrestrial planet, making such a measurement easy for an infrared direct imaging mission \citep[e.g., NASA's Terrestrial Planet Finder Interferometer, TPF-I, or ESA's Darwin;][respectively]{Beichman_1998, Fridlund_2000}.\footnote{It is worth noting, however, that the albedo of snow, ice and clouds is much less for planets orbiting low-mass stars \citep{Joshi_2011}.  This makes it harder for such planets to enter a runaway snowball, but also makes it harder to recognize such snowballs, once formed.}  However, an albedo estimate is necessary, but insufficient, for understanding a planet's climate. Perversely, the end state at the inner edge of the HZ is also highly reflective, if Venus is any indication. 

A climate constraint orthogonal to albedo is planetary thermal inertia, which depends on atmospheric and surface characteristics. Terrestrial planets may be divided into four categories based on albedo and thermal inertia: 1) low albedo and low thermal inertia (Mars), 2) high albedo and low thermal inertia (snowball), 3) high albedo and high thermal inertia (Venus), and 4) low albedo and high thermal inertia (Earth). Furthermore, a large thermal inertia for a planet with a modest atmosphere would constitute circumstantial evidence for liquid water oceans. 

A planet's  thermal inertia can be estimated based on its response to external radiative forcing.  \cite{Gaidos_2004} used an analytic energy balance model (EBM), which assumes diffusive heat transport, to explore how the thermal phase variations of terrestrial planets are affected by a planet's eccentricity, obliquity and thermal inertia. They found strong degeneracies between these factors: even if the planet's orbit is well characterized via astrometry or imaging, the obliquity, equinox phase and thermal inertia cannot be uniquely determined without detecting second-order phase variations.  

The present study goes beyond the important work of \cite{Gaidos_2004} in the following ways: 1) we use a GCM rather than an analytic EBM, allowing us to self-consistently model the three-dimensional atmosphere and ocean dynamics of these planets, 2) we test the effect of planetary thermal inertia using temperate and snowball climates, which are realistic and stable on geological timescales, 3) we include the effects of diurnal heating and rotational variability, and 4) we simulate ten years worth of observations for each planet, allowing us to evaluate the repeatability of these patterns given stochastic weather and inter-annual variability. Finally, our study is complementary to \cite{Gaidos_2004}: we are unable to vary eccentricity or obliquity because of computational expense, but we consider a wide variety of viewing geometries.

\subsection{Viewing Geometry}\label{viewing_geometry}
\begin{figure}[htb]
     \vspace{-1.0cm}
      \includegraphics[angle=270, width=80mm]{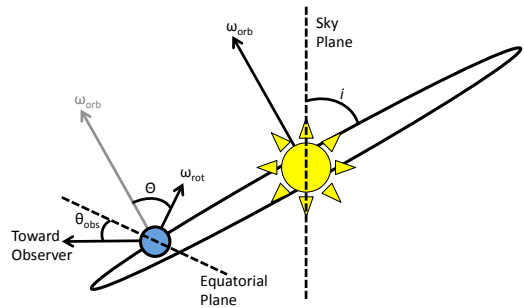}
    \vspace{-1.0cm}
    \caption{{\small The viewing geometry is defined by any three of $i$, $\theta_{\rm obs}$, $\Theta$, or $\xi_{\Theta}$. Orbital inclination, $i$ is the angle between the orbital plane of the planet and the sky plane, or equivalently the angle between the orbital angular momentum and the line of sight.  The sub-observer latitude, $\theta_{\rm obs}$, is the latitude directly below the observer.  These two angles are related via the obliquity, $\Theta$, and equinox phase, $\xi_{\Theta}$, via Equation~\ref{lat_obs}. \label{geometry}}}
\end{figure}

It is useful to first define the system geometry, shown in Figure~\ref{geometry}.  An exoplanet's orbital inclination, $i$, denotes how far it is from being in a face-on orbit: a planet with $i=0$ orbits in the plane of the sky, while $i=90^\circ$ means an edge-on orbit. Unless a planet is in a perfectly face-on orbit, it exhibits phases: when the planet is closest to us, inferior conjunction, we primarily see the planet's nightside; when the planet is farthest from us, superior conjunction, we see mostly dayside. A planet on an edge-on orbit is ``new'' at inferior conjunction, and ``full'' at superior conjunction. Quadrature occurs when we see half of the illuminated hemisphere of the planet (as with the Moon's quarter phases). A planet on a face-on orbit is permanently in quadrature; for other inclinations, quadrature occurs twice per orbit.

The angle between the planet's rotation axis, $\vec{\omega}_{\rm rot}$, and its orbital axis, $\vec{\omega}_{\rm orb}$ is the obliquity, $\Theta$. The orientation of the rotational axis with respect to an observer is specified by the equinox phase, $\xi_{\Theta}$. Note that equinox phase is related to ---but not the same as--- precession phase, the orientation of the planet's spin axis with respect to periastron. 

The sub-observer latitude, $\theta_{\rm obs}$, is the latitude of the point directly beneath the observer.  If the planet's equator is directly beneath the observer, then $\theta_{\rm obs}=0^\circ$ (``equator-on'' viewing geometry), while $\theta_{\rm obs}=90^\circ$ indicates that the observer is located directly above the planet's north pole (``pole-on'' viewing geometry).

The sub-observer latitude is related to the inclination, planetary obliquity,  and equinox phase:
\begin{equation}\label{lat_obs}
\sin\theta_{\rm obs} = \sin\Theta\cos\xi_{\Theta}\sin i + \cos\Theta\cos i,
\end{equation}
where conservation of angular momentum dictates that all of the parameters in the equation are constants, provided that we neglect precession.

\subsection{Sources of Thermal Phase Variations}
\begin{figure*}[htb]
     \vspace{-2.5cm}
      \includegraphics[angle=270, width=160mm]{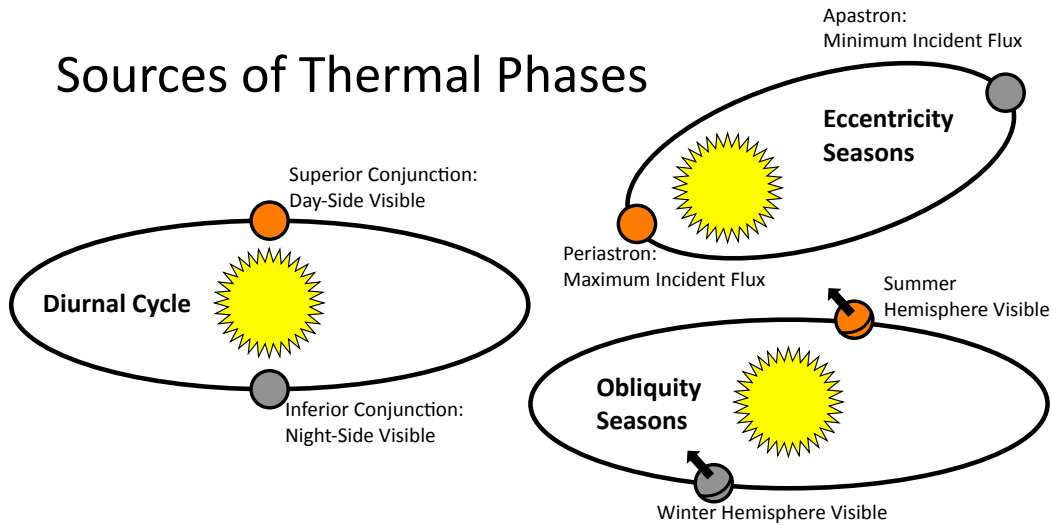}
    \vspace{-2.5cm}
    \caption{{\small The three periodic drivers of climate, the diurnal cycle, obliquity seasons, and global seasons, are the sources of thermal phase variations.  The response of a planet to these drivers depends on its thermal inertia and heat transport, as described in the text and in Table~\ref{thermal_phases_table}. \label{thermal_phases}}}
\end{figure*}

An externally heated planetary atmosphere can roughly be thought of as a heat engine. For example, surface-to-tropopause and equator-to-pole heating gradients on Earth lead to steady-state transport of energy: upward and poleward, respectively.  It is the higher-order periodic climate forcings, however, that cause thermal phase variations (Table~\ref{thermal_phases_table} and Figure~\ref{thermal_phases}). These variations are damped by heat storage (i.e., thermal inertia) and transport (atmospheric and/or oceanic). 

\begin{deluxetable*}{lllll}
\tabletypesize{\scriptsize}
\tablecaption{External Periodic Climate Forcing and the Sources of Thermal Phases \label{thermal_phases_table}}
\tablewidth{0pt}
\tablehead{
\colhead{Climate Forcing} & \colhead{Frequency} & \colhead{Amplitude$^a$} & \colhead{Damping Mechanism}&\colhead{Favored Geometry}}
\startdata
Diurnal Cycle & Rotational$^b$ & $\propto S_0 (1-A)$ & Heat Storage and Zonal Transport$^c$ & Edge-On Orbit\\
Eccentricity Seasons & Orbital & $\propto \frac{4eS_0(1-A)}{1-2e^2 - 2e^3 + e^4}$ & Heat Storage & Always Present\\
Obliquity Seasons & Orbital & $\propto S_0 (1-A)\sin\Theta$ & Heat Storage and Meridional Transport & Pole-On Rotation 
\enddata
\tablenotetext{a}{The intrinsic peak-to-trough forcing amplitude, $S_{\rm max}-S_{\rm min}$, is expressed in terms of the Solar constant, $S_0$, the planet's Bond albedo, $A$, orbital eccentricity, $e$, and obliquity, $\Theta$. The more familiar relative amplitude for eccentricity seasons is $S_{\rm max}/S_{\rm min} = \left((1+e)/(1-e)\right)^2$.  The $\sin\Theta$ scaling for obliquity seasons is approximately correct for mid--high latitudes and thus for the hemispherical average \citep[we determined this numerically solving the expressions in Section 7.3 of][]{Pierrehumbert_book}.}
\tablenotetext{b}{The diurnal \emph{forcing} occurs at the rotational frequency, but it manifests itself to a distant observer on the orbital timescale as phase variations.}
\tablenotetext{c}{On Earth, wind speeds are much slower than rotational velocity for all but polar regions, so the diurnal cycle is almost entirely damped by thermal inertia.}
\end{deluxetable*}

\textit{Diurnal Cycle:} A planet receives flux from its host star during the day but not at night.  This tends to make a planet's day-side warmer than its night side, leading to phase variations: the day-side is most visible near superior conjunction, and the night side is most visible near inferior conjunction. Insofar as a planet has some thermal inertia, the day--night temperature contrast is reduced and the hottest time of day occurs in the afternoon rather than at noon.  If the advective timescale is comparable to, or shorter than, the rotational period, then zonal (E--W) heat transport becomes an important factor for damping the diurnal temperature swings. The diurnal cycle is most observable for planets on edge-on orbits, and is the sole source of phase variations for a zero-eccentricity, zero-obliquity planet. 

If a planet's surface has inhomogeneous emitting temperature (e.g., due to patchy clouds), then different regions of the planet will have different emitting temperatures at the same local solar time, leading to \emph{rotational} variability. This manifests itself to an external, inertial observer as high-frequency variations in emitting temperature, where the magnitude of variations depends ultimately on the longitudinal albedo and/or thermal inertia contrasts on the planet. While this effect is included in our simulations and could in principle be used to determine a planet's rotation rate \citep{Gomez-Leal_2012}, it would be averaged over for realistic integration times ($> 24$~hrs), and we do not discuss it further in this paper.

\textit{Eccentricity Seasons:} A planet with non-zero orbital eccentricity tends to be hottest near periastron and coolest near apastron.  If the planet has some thermal inertia, the seasonal temperature swings are damped, and the maximum temperature occurs somewhat after periastron. Orbital eccentricity affects the entire planet's power budget,  and hence affects thermal phase variations seen from any vantage point; eccentricity effects may be isolated if a zero-obliquity planet is viewed face-on.

\textit{Obliquity Seasons:}  Non-zero obliquity leads to the northern hemisphere being hotter during northern summer and cooler during northern winter. Obliquity seasons are damped by thermal inertia and meridional (N--S) heat transport.  Insofar as one hemisphere is more visible to the observer ($\theta_{\rm obs} \ne 0$), obliquity seasons cause thermal phase variations. If the northern and southern hemispheres have very different albedos and/or thermal inertia, the planet's global power budget will exhibit seasonal variations regardless of viewing geometry, much as with eccentricity seasons. In general, obliquity seasons are most extreme as seen by a pole-on observer. A planet on a circular, face-on orbit will have thermal variations entirely dictated by obliquity seasons.

\subsection{Snowball Climate}
A terrestrial planet in the HZ of its host star \citep{Hart_1979, Abe_1993, Kasting_1993} does not necessarily have a temperate climate or harbor liquid water on its surface.  The high reflectivity of ice and snow compared to land and ocean allows an Earth-like planet at the exact same distance from its star
and with the same atmosphere to exist in either a temperate
state, like modern Earth, or an icy snowball state \citep{Budyko_1969, Sellers_1969, Marotzke_2007}. 

It is believed, for example, that Earth has on
occasion ($\sim$710~Ma, $\sim$635~Ma, and possibly at other times) fallen into a snowball state, during which global oceans completely \citep[][]{Kirschvink_1992, Hoffman_1998} or
nearly completely \citep[][]{Hyde_2000,Abbot_2011b} froze
over. As a limiting case, we
will be considering here a planet with entirely ice-covered oceans. The
proximate mechanism for the initiation of a snowball state is a
runaway ice/snow-albedo feedback: replacing ocean with sea ice
leads to cooling and more ice formation \citep{Budyko_1969, Sellers_1969}.
The ultimate cause of the specific snowball events in Earth's history
is still unknown \citep{Pierrehumbert_2011b}.

The high albedo of ice and snow make a snowball state, once formed, extremely stable \citep{Budyko_1969, Sellers_1969, Marotzke_2007}.  Compared to Earth's 4.5~Gyr geological history, snowball events have been rare and brief (millions of years), with a duty cycle of order 1\%, probably due to the geologically rapid silicate-weathering feedback \citep[][]{Walker_1981}. The globally frozen continents and low temperatures halted silicate weathering and the associated pulldown of atmospheric CO$_{2}$, but greenhouse gases kept being resupplied by volcanism and plate tectonics.  The resulting increase in greenhouse warming ---in conjunction with other effects such as the dirtying of tropical sea ice \citep{Abbot_2010}--- snapped Earth out of the snowball state. 

It is not known how the escape from a snowball state would play out on exoplanets. Possible differences include the presence of a large water reservoir \citep[wet planets are more prone to runaway snowball than ``Dune'' planets;][]{Abe_2011}, incident stellar flux \citep[reduced power budgets make planets more susceptible to global glaciations, including the limit of a starless planet,][]{Tajika_2008, Abbot_2011}, host star spectral type \citep[bluer stars lead to a stronger temperature--albedo feedback;][]{Joshi_2011}, planetary obliquity \citep[low-obliquity planets are slightly more prone to snowball events:][]{Williams_1997, Williams_2003, Spiegel_2009}, orbital eccentricity \citep[larger eccentricity slightly increases the propensity of a planet to enter a snowball state:][]{Williams_2002, Dressing_2010}, planetary rotation and size (slower rotation or a smaller planet increases meridional transport, making it easier to enter a snowball state), and the effectiveness of an extrasolar carbon-silicate cycle (continent configuration, water levels, nature of mantle convection, etc.). 
  
Most importantly for our purposes, a snowball climate has relatively little surface thermal inertia compared to a temperate climate: a few meters equivalent depth of water compared to the current temperate climate with its tens of meters equivalent depth of water.\footnote{Earth's oceans are on average 3~km deep, but only the top tens of meters participate in damping seasonal variations.}  Trying to distinguish between these climates based solely on thermal phase variations is therefore a good test for the utility of thermal phase variations in constraining the heat capacity of HZ terrestrial exoplanets.

\section{Global Climate Models}
To simulate thermal phase variations for a variety of viewing geometries,
we use output from the coupled ocean-atmosphere global 
climate model ECHAM5/MPI-OM. A detailed description
of the GCM is given in \cite{Voigt_2011} and references
therein. We use top-of-atmosphere (TOA) incident stellar flux, outgoing longwave radiation, and reflected light, each of which is outputted every two hours at each grid point.

\begin{figure*}[htb]
\begin{center}$
\begin{array}{cc}
\includegraphics[width=84mm]{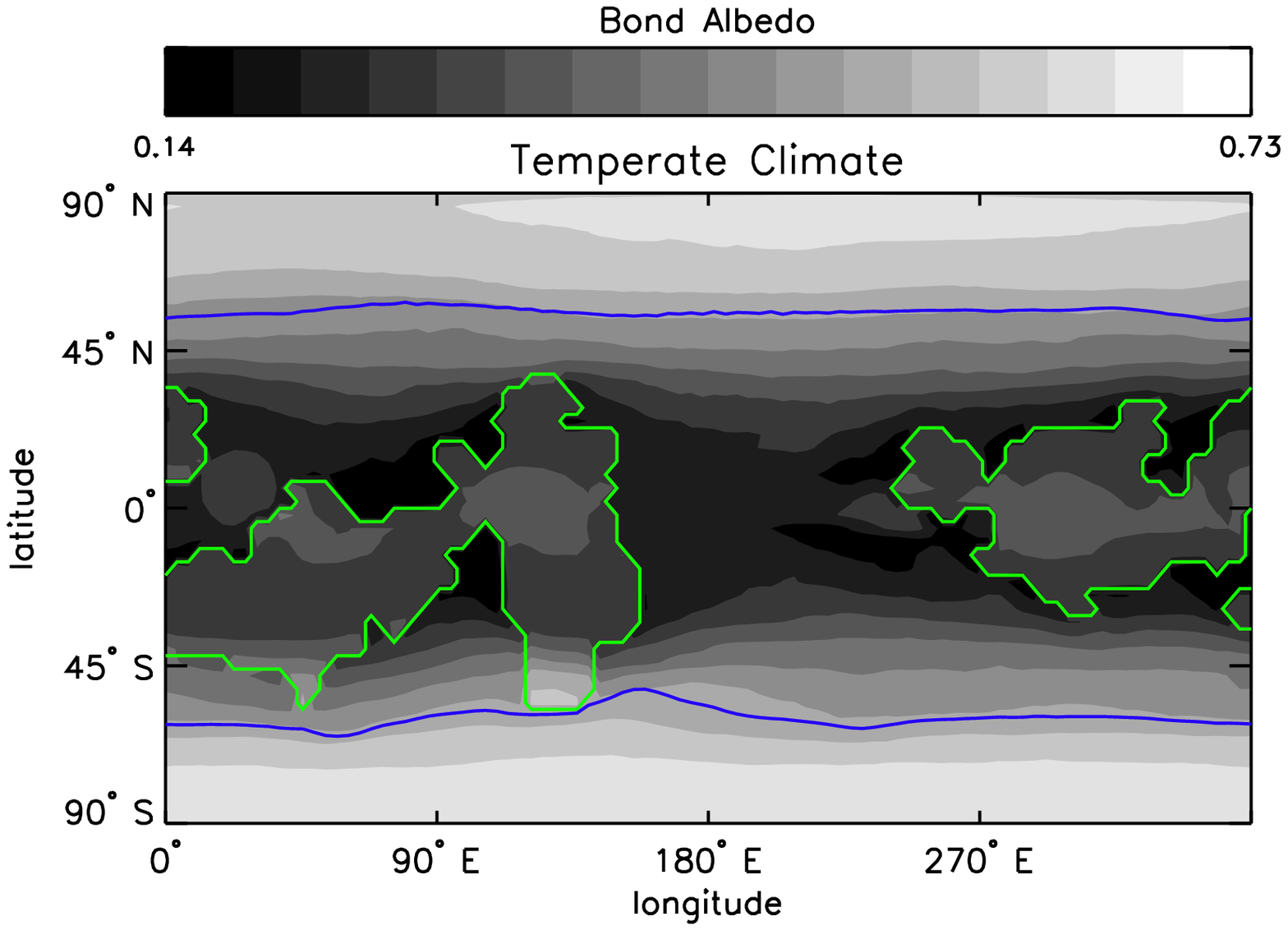} & \includegraphics[width=84mm]{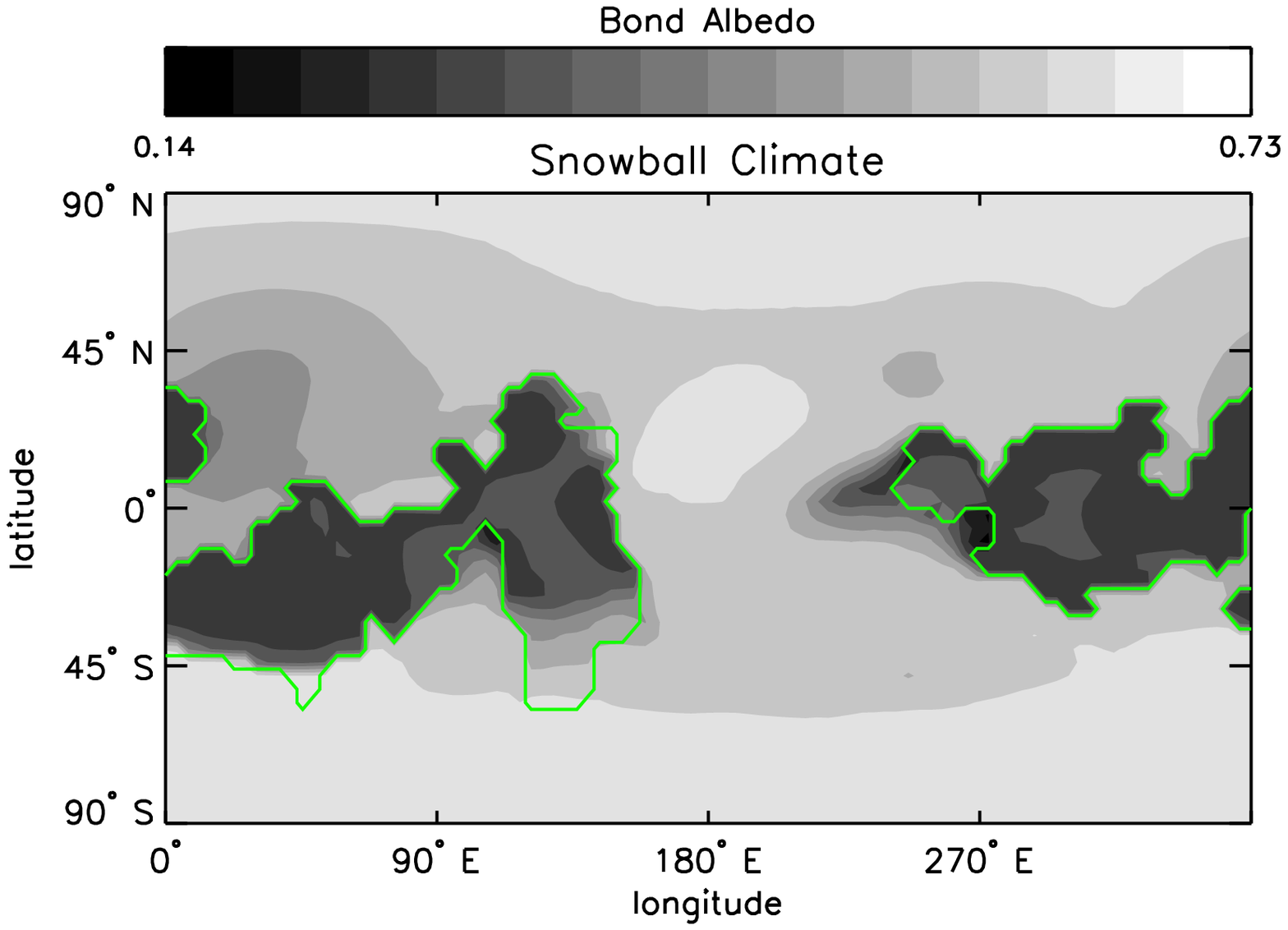}\\
\includegraphics[width=84mm]{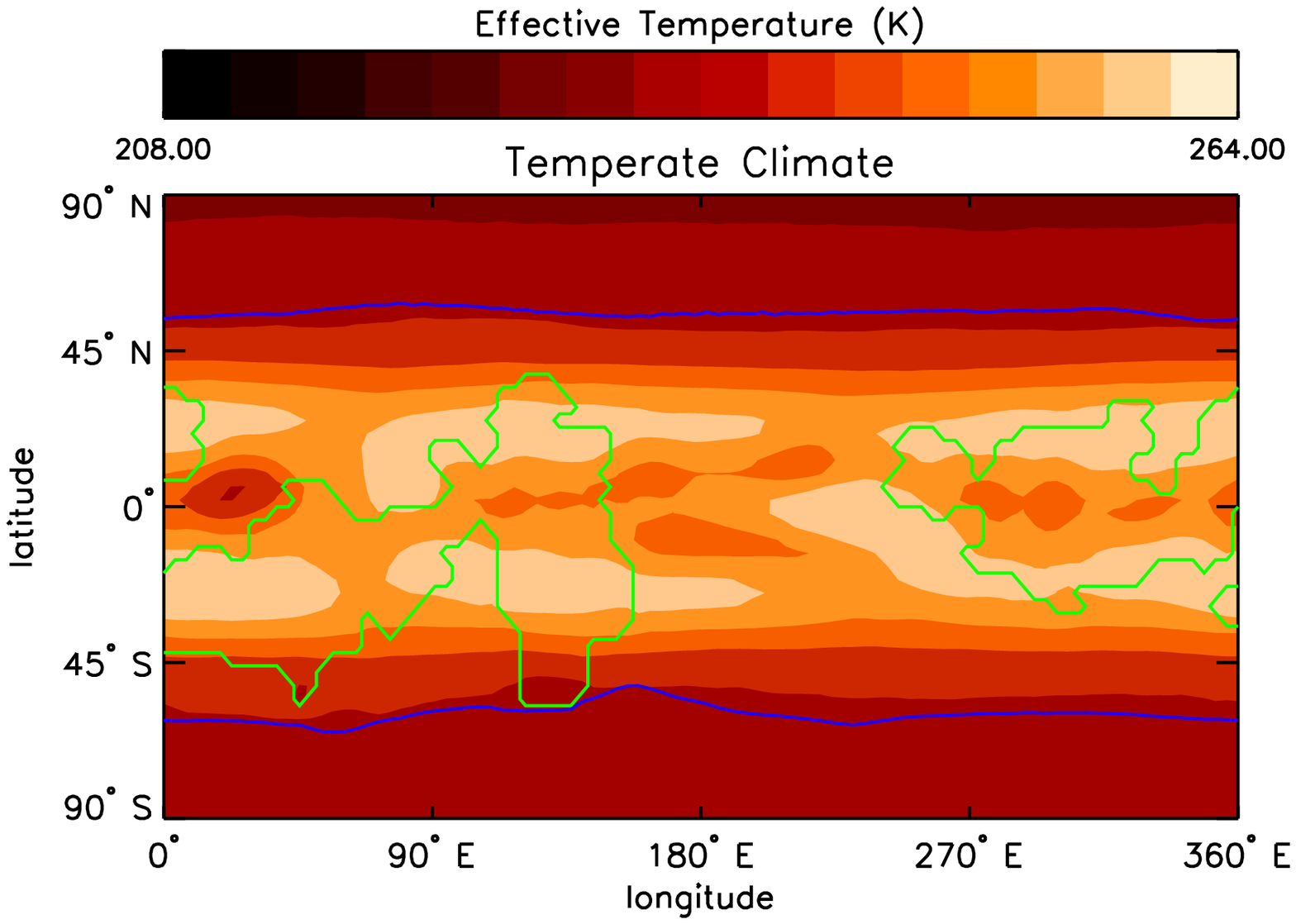} & \includegraphics[width=84mm]{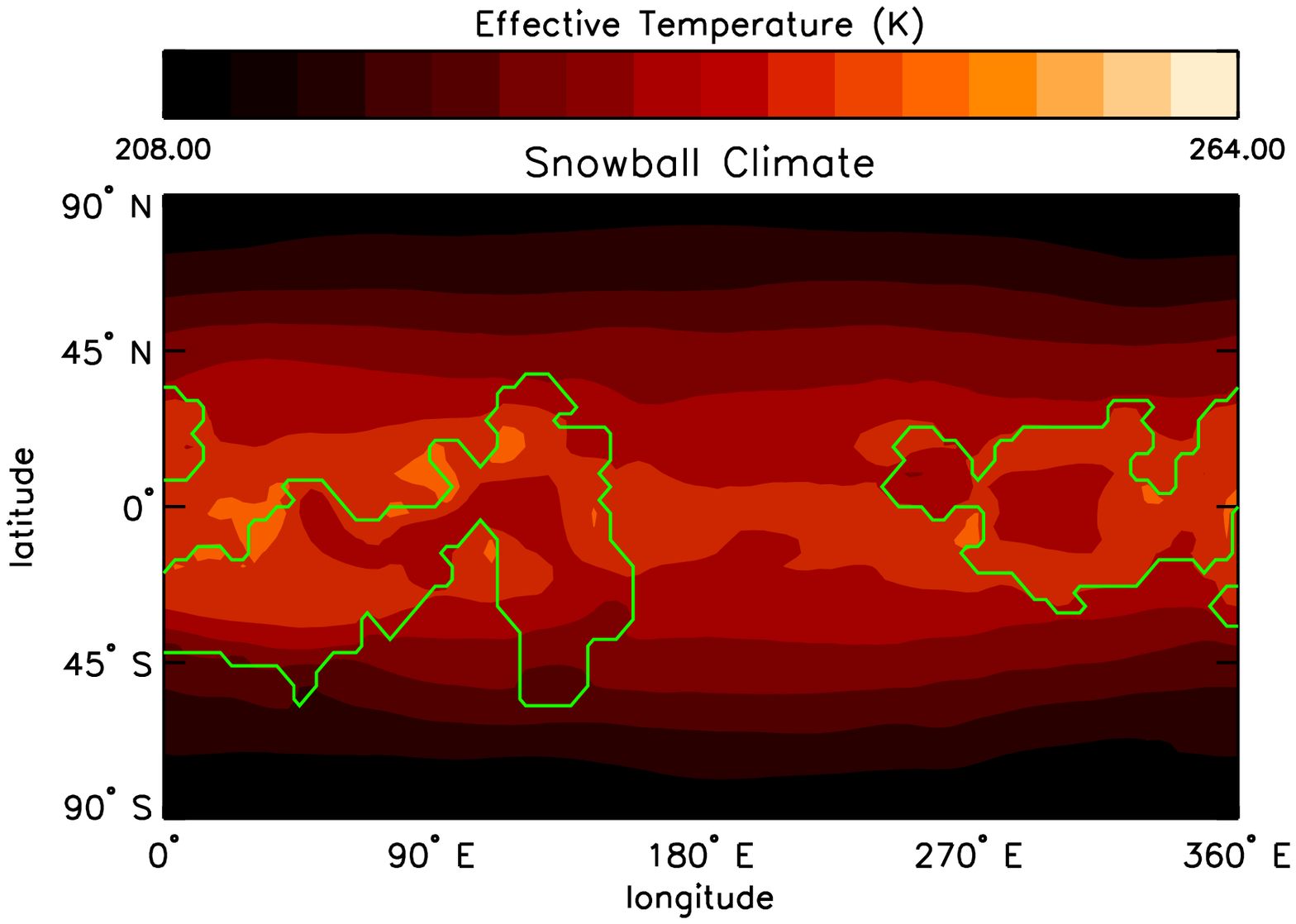}
\end{array}$ 
\end{center}
\caption{Ten-year averaged maps of top-of-atmosphere Bond albedo (top panels) and effective temperature (bottom panels) for the temperate (left) and snowball (right) models. Green lines indicate the coasts of the equatorial continents; the blue lines show the annual mean extent of sea ice in the temperate model.}
\label{average_maps}
\end{figure*}

Specifically, 
we here use two simulations presented in
\cite{Voigt_2011}. These simulations use  
a modern-Earth solar constant of 1367\,Wm$^{-2}$, pre-industrial
atmospheric carbon dioxide (278~ppm), a modern-Earth
orbit (23.5$^\circ$ obliquity, 1.7\% eccentricity) and two large tropical supercontinents that are meant to
represent Earth's topography 635 Ma before present. The time-averaged albedo and temperature maps of the two models are shown in Figure~\ref{average_maps}. 

The ``temperate''
simulation is initiated from a climate state with polar ice caps and remains in
such a state throughout.  The ``snowball'' simulation is initiated from a climate state with globally ice-covered oceans and remains in such a state throughout. As found in previous snowball climate simulations, the low latitudes, and in particular the majority of the continental area, remain bare because of a reduced hydrological cycle and net evaporation \citep[e.g.,][]{Pierrehumbert_2005}. Both simulations were run to equilibrium, then for an additional ten
years, with shortwave and longwave radiation outputs saved every two hours.

Although our temperate simulation is roughly commensurate with the modern-day Earth, it differs in two important ways: 1) the continental configuration is much more N--S symmetric, reducing the impact of obliquity on the global power budget, and 2) the continents are more uniform longitudinally, reducing the amplitude of rotational variations \citep[cf.][]{Gomez-Leal_2012}.  These differences make the resulting climate variations of our temperate planet, if anything, more straightforward to interpret than the modern Earth, strengthening our ultimate conclusions.

\section{Time-Averaged Power Budget}
The time-averaged, total absorbed power of the planet must equal its emitted thermal power. We verify that this is indeed the case, to one part in a thousand, for the temperate simulation, while the snowball's emitted flux exceeds its absorbed flux by 5.5\%. More energy leaves the atmosphere at the top than enters at the bottom, yet the snowball planet does not cool off, implying that there is a small spurious heat source in the model atmosphere. This discrepancy, which may be a surface diagnostic bug rather than a true model error, was noted by \cite{Voigt_2011} and should not significantly affect our results, since we only compare \emph{time variations} in thermal emission. We simply renormalize the snowball simulation's thermal emission to the time-averaged absorbed power in Sections 3 \& 4 of this paper.

\subsection{Year-Averaged Meridional Heat Transport}
\begin{figure*}[htb]
\begin{center}$
\begin{array}{cc}
\includegraphics[width=84mm]{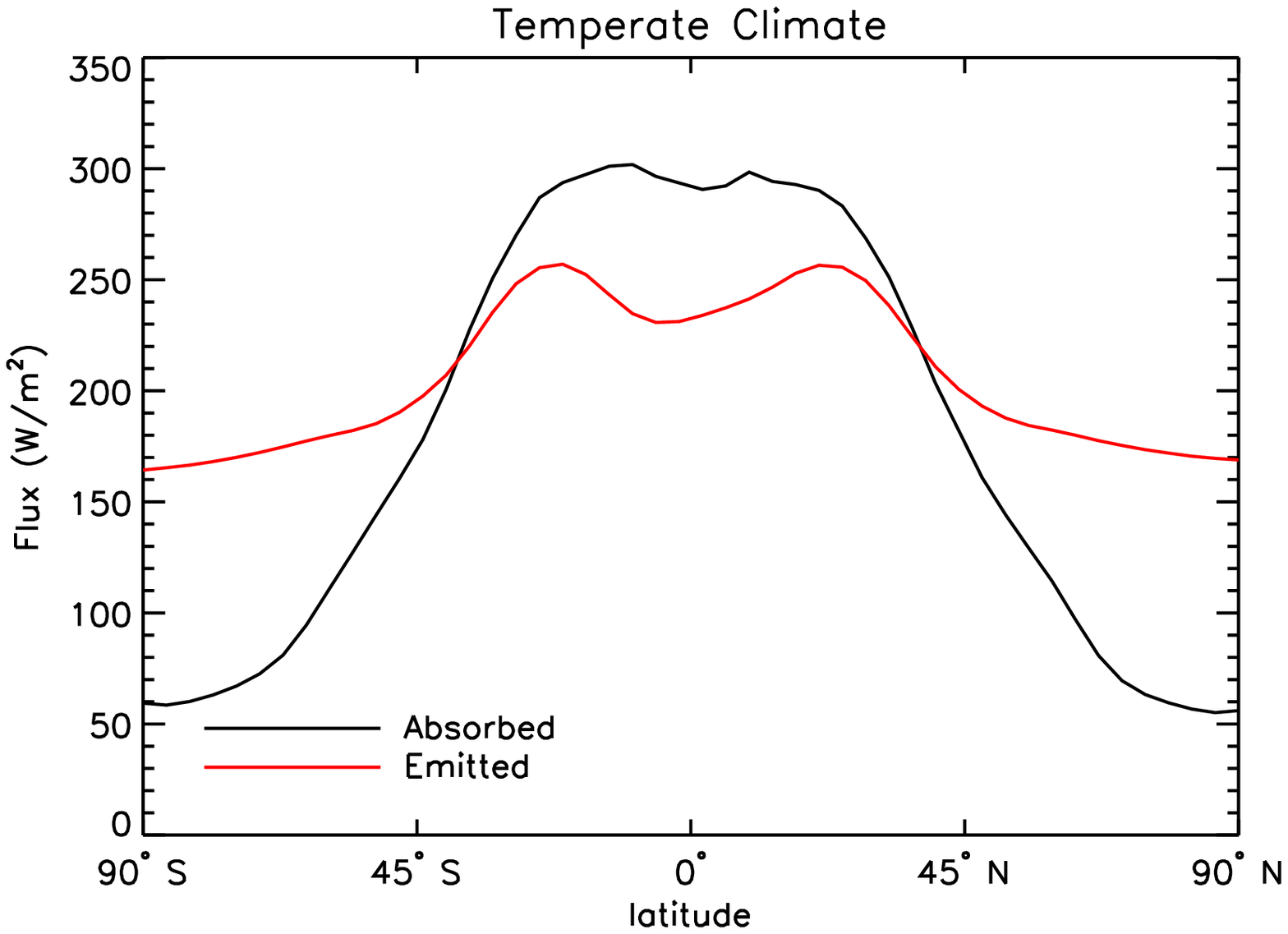} & \includegraphics[width=84mm]{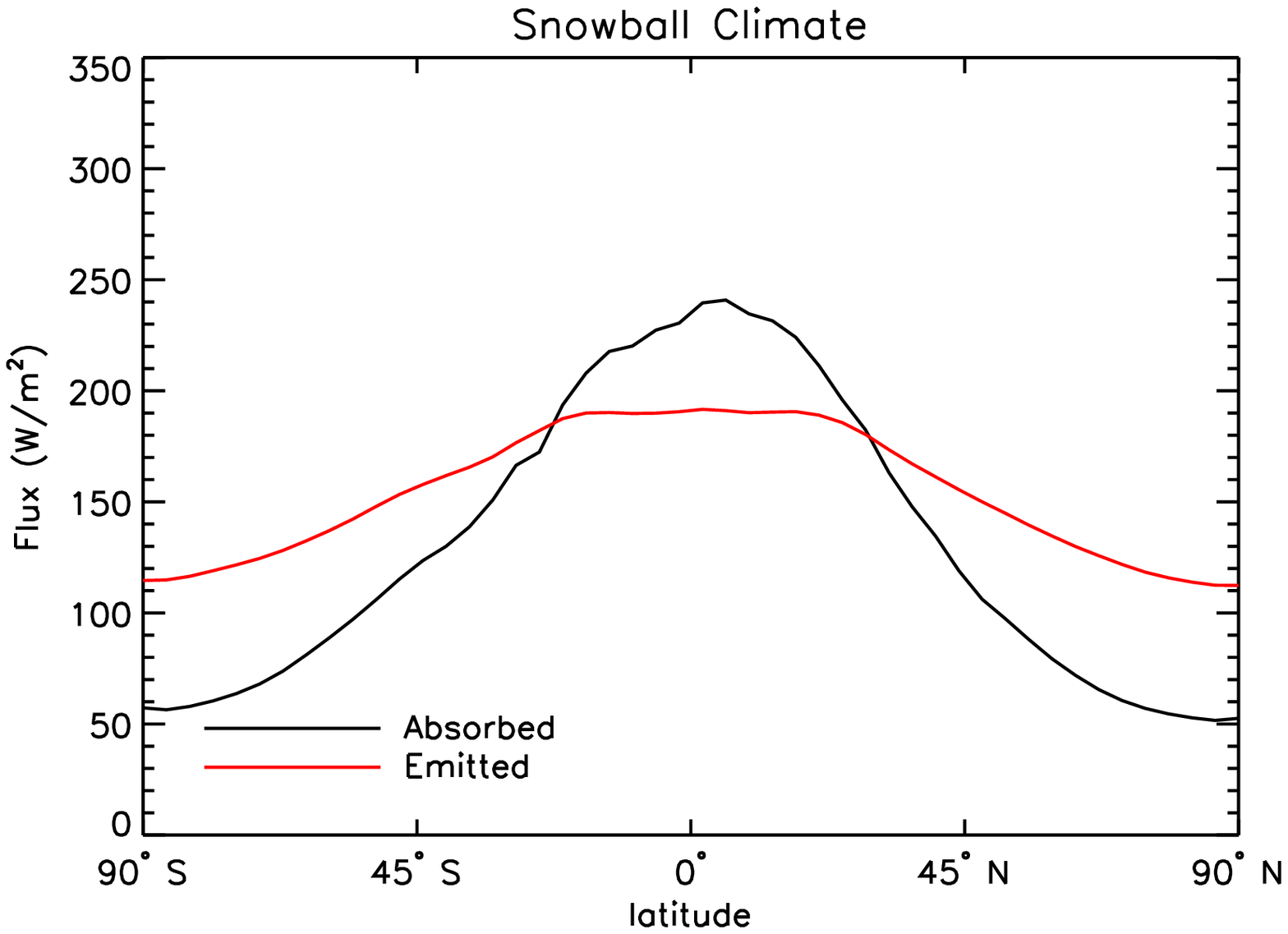}
\end{array}$ 
\end{center}
\caption{Top-of-atmosphere absorbed (black) and emitted (red) flux as a function of latitude.}
\label{toa_radiative_budget}
\end{figure*}

The modest obliquity of our simulated planets dictates that high latitudes receive less orbit-averaged insolation than low latitudes.  In Figure~\ref{toa_radiative_budget} we show the top-of-atmosphere radiative budget for the two simulations.  Averaging over the seasonal variations that will be the subject of the bulk of this paper, the equator--pole gradient in absorbed flux (black lines) results in a meridional temperature gradient, which drives poleward energy transport. As a result, the equator--pole gradient in thermal emission (red lines) is muted compared to the gradient in absorbed flux. 

Comparing the radiative budgets of the two simulations, it is clear that the meridional radiative forcing gradient is somewhat greater for the temperate case than for the snowball (240 vs. 200 W/m$^2$), while the resulting gradient in thermal emission is similar for the two planets (80 vs. 70 W/m$^2$). This indicates that more poleward heat transport is occurring on the temperate planet.    

\begin{figure}[htb]
\begin{center}
\includegraphics[width=84mm]{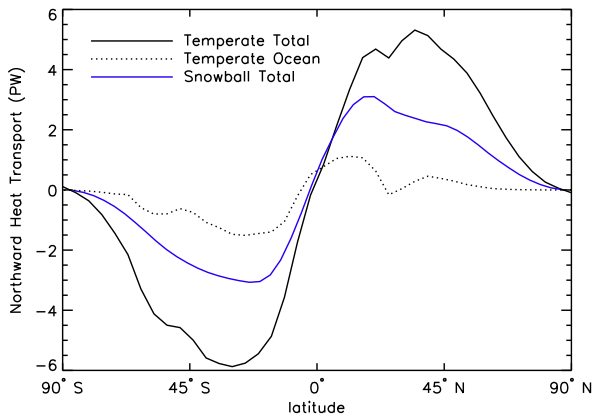}
\end{center}
\caption{Seasonally-averaged northward heat transport for the temperate (black) and snowball (blue) climates. The snowball planet has ice-covered oceans and thus negligible oceanic heat transport.}
\label{meridional_transport}
\end{figure}

Figure~\ref{meridional_transport} shows the orbit-averaged northward heat transport for the two simulations; heat transport on the temperate planet is roughly twice that on the snowball. Some of this difference can be attributed to the weaker meridional forcing on the snowball planet.  Moreover, transport is much greater in the temperate climate because of two compounding effects: the presence of ocean transport and greater atmospheric humidity.

On their own, neither the absence of ocean heat transport, nor the drier air in principle need to grossly affect the total meridional heat transport of an Earth-like planet due to  ``compensations'' noted and explained by \cite{Stone_1978}, \cite{Vallis_2009}, and \cite{Enderton_2009}.  But in combination, these two changes make it impossible for the climate to fully compensate, with the net effect that the mid-latitude meridional heat transport in the snowball simulation is significantly reduced. 

\section{Periodic Climate Forcing}\label{climate_forcing}
We now consider periodic changes in the planet's power budget. We first develop a simple model for interpreting time-variations in absorbed and emitted power, then consider each of the periodic climate forcings in turn, as they manifest themselves in our simulations.  The planet's response to radiative forcing is independent of an external observer, but is the ultimate cause of the thermal phase variations discussed in Section~5.

\subsection{Response to Periodic Forcing}
In order to quantitatively estimate thermal inertia from periodic variations in absorbed and emitted power, we approximate the climate system with the following differential equation \citep[e.g.,][]{Pierrehumbert_book}:
\begin{equation} \label{diff_equ}
\mu \frac{d T}{d t} = F_{0} + \Delta F_{\rm abs} \cos (\omega t) - bT,
\end{equation}
where $\mu$ is the effective thermal inertia coefficient, $T$ is the emitting temperature, $\Delta F_{\rm abs}$ is the semi-amplitude of variations in the absorbed power, $\omega$ is the driving frequency, and $b$ linearly approximates the relation between temperature and emitted flux \citep[the exact differential equation, analytic and numerical solutions are developed for a blackbody in][]{Cowan_2011a}.  

It is worth noting that this formalism is usually adopted for \emph{surface} temperature variations.  Nonetheless, the most accessible temperature for any planet other than Earth is the \emph{emitting} temperature so it is natural to consider effective thermal inertia as it affects top-of-atmosphere absorbed and emitted flux.  Inferring bulk thermal inertia for the underlying atmosphere and surface is of course complicated when one is only sensitive to the emitting temperature.  For example, the role of greenhouse gases in Equation~\ref{diff_equ} is normally to reduce $b$, since for a given change in surface temperature, less flux escapes to space.  In the current scenario where $T$ denotes top-of-atmosphere emitting temperature, the emitted flux is $\sigma T^4$, where $\sigma$ is the Stefan-Boltzmann constant. In the limit of small temperature excursions, we may adopt $b=4\sigma T_{0}^3$, where $T_{0}$ is the average emitting temperature. The greenhouse effect is therefore folded into $\mu$: greater infrared opacity increases the atmosphere's radiative timescale and hence the lag time between radiative forcing at the surface and response at the top-of-atmosphere. Insofar as upward energy transport in the troposphere is primarily accomplished via convection, the atmosphere's infrared opacity determines the pressure of the emitting layer. Either way, the net result of increasing the concentration of infrared absorbers is increasing the effective thermal inertia coefficient. 

The periodic solution to Equation~\ref{diff_equ} is 
\begin{equation}
T(t) = \frac{F_{0}}{b} + \Delta T \cos (\omega t - \Phi),
\end{equation}
where the temperature oscillations have a semi-amplitude $\Delta T = \Delta F_{\rm abs}/(b\sqrt{1+\omega^2 \tau^2})$, the phase offset is given by $\tan\Phi=\omega\tau$, and the thermal relaxation time is $\tau \equiv \mu/b$.
 
In this linearized form, the emitted flux is proportional to temperature, $\Delta F_{\rm emi} = b \Delta T$, so one can estimate the radiative relaxation time based on the amplitudes of variations in absorbed and emitted power: 
\begin{equation}\label{tau}
\tau = \frac{1}{\omega}\sqrt{\left(\frac{\Delta F_{\rm abs}}{\Delta F_{\rm emi}}\right)^2 -1}. 
\end{equation}

\subsection{Global-Mean Temperature Variations} \label{eccentricity_section}
The incoming shortwave radiation undergoes a 6.8\% peak-to-trough orbital modulation because of Earth's 1.7\% eccentricity, while the albedo exhibits seasonal variations of 6\% and 4\% in the temperate and snowball climates, respectively. The albedo fluctuations are primarily driven by meridional (N--S) differences in albedo coupled with seasonal variations in the sub-solar latitude \citep[][]{Cowan_2012b}. Seasonal changes in cloud cover may also contribute to the changing albedo. This effect cannot be predicted \emph{a priori} but is an emergent phenomenon of the GCM.

\begin{figure*}[htb]
\begin{center}$
\begin{array}{cc}
\includegraphics[width=84mm]{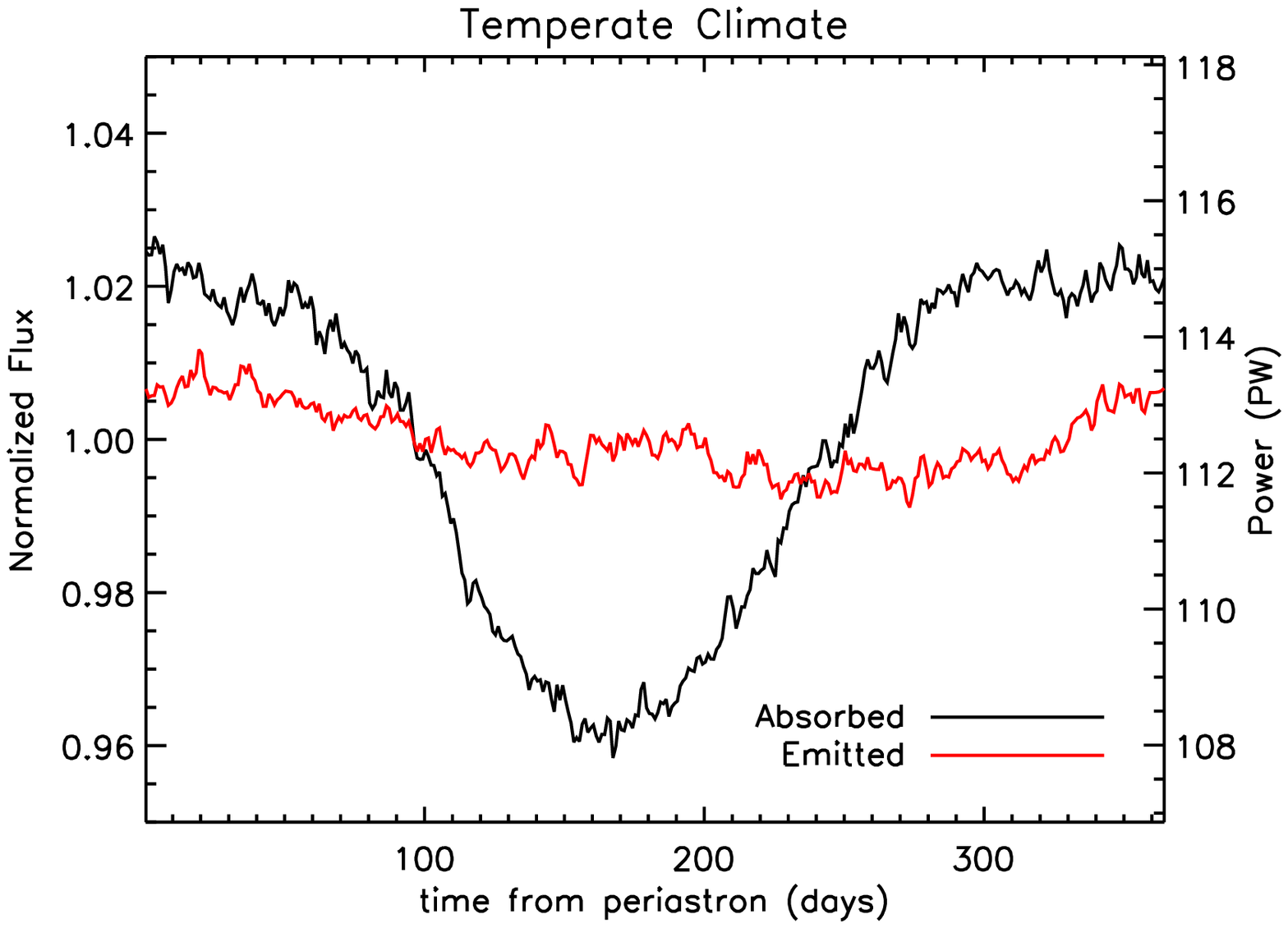} & \includegraphics[width=84mm]{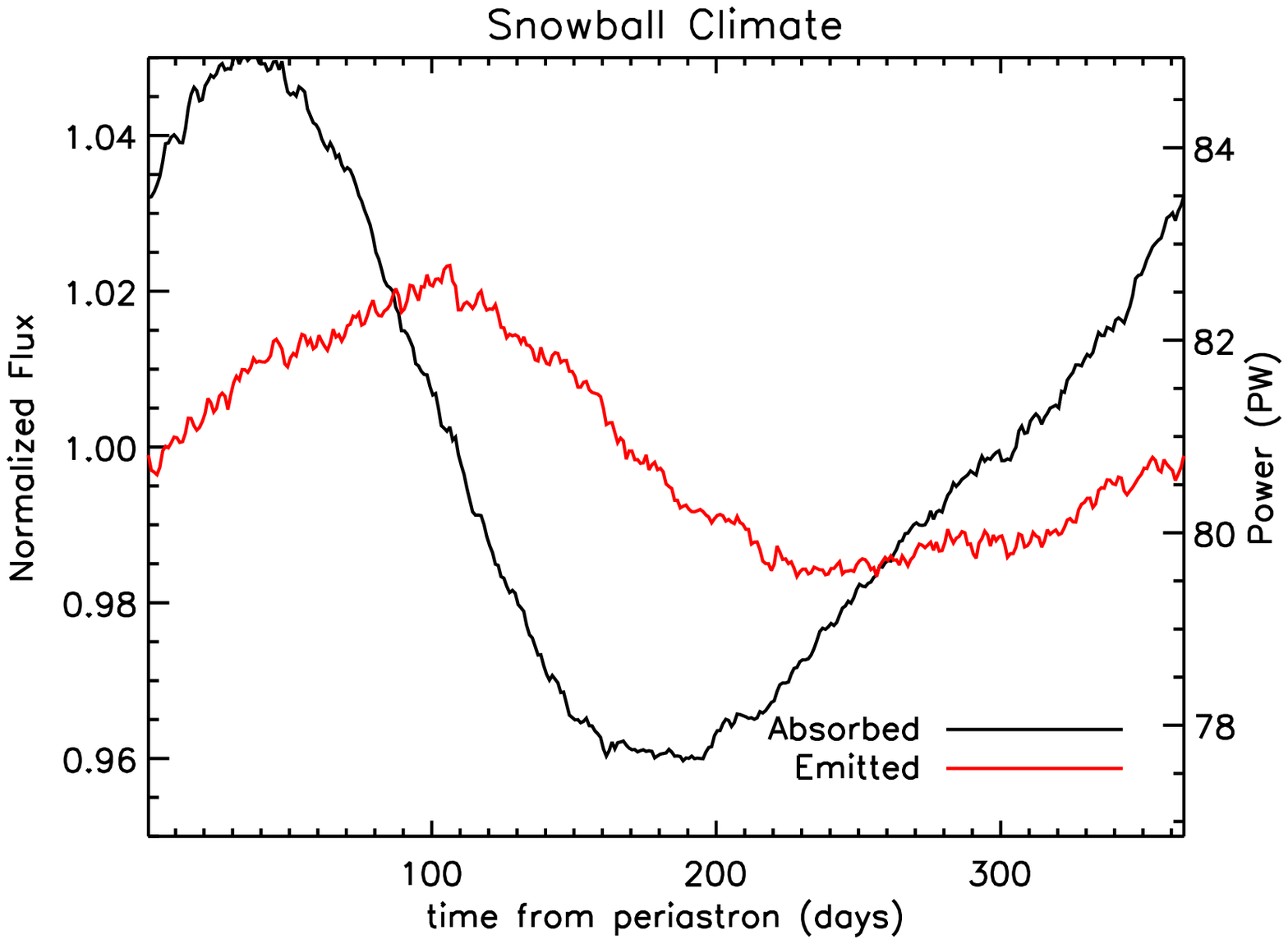}
\end{array}$ 
\end{center}
\caption{Planet-integrated, top of atmosphere fluxes for a temperate climate (left) and a snowball climate (right). The plots were made by combining  ten years of simulations and averaging over rotational variability. This plot, and most others in the paper, uses the same relative flux scaling (left axes) but not the same absolute scaling (right axes) for the temperate and snowball simulations.}
\label{planet_averaged}
\end{figure*}

The combination of changing incident flux at the top of the atmosphere and changing albedo leads to peak--trough variations in absorbed power of 6\% and 8\% for the temperate and snowball climates, shown with black lines in Figure~\ref{planet_averaged} (in physical units, the two simulations experience approximately the same 7~PW forcing).\footnote{The peak-to-trough amplitudes are determined by fitting a sinusoid to the absorbed and emitted curves.} The changes in absorbed flux drive ``global'' seasons.  The critical difference between global seasons and the usual seasons we experience on Earth are that the seasons are in phase in the northern and southern hemispheres, so cross-equatorial heat transport cannot damp them.  Insofar as the typical eccentricity of exoplanets is 0.1--0.3 \citep{Ford_2008, Moorhead_2011}, it is possible that eccentricity-driven global seasons will be the norm on other planets. 

The thermal inertia of the planet damps the amplitude of the oscillations and produces a phase lag, as seen in the planet-integrated emitted power, shown in red in Figure~\ref{planet_averaged} (peak--trough seasonal changes of 1\% and 3\% for the temperate and snowball models, respectively).  Note that the low-amplitude response of the temperate planet makes the phase offset difficult to quantify. For the temperate planet, peak emission occurs nearly simultaneously with the peak absorption, but the minimum emission lags the minimum in forcing by many months. 

Using Equation~\ref{tau} we obtain $\tau = 343$~days and 145~days for the temperate and snowball climates, respectively. The temperate planet therefore has a thermal relaxation time 2.4$\times$ longer than that for the snowball. Note that these are longer than the \emph{atmospheric} thermal relaxation times because we are accounting for surface heat capacity. 

It is often convenient to express the thermal inertia coefficient, $\mu$, as an equivalent mixed layer depth (MLD) of water, since oceans are the dominant source of thermal inertia on Earth: $H = \mu/(\rho c_p)$, where $\rho$ and $c_p$ are the density and specific heat capacity of water. Using the globally-averaged emitting temperatures of the two simulations ($T_0 = 249$~K and 232~K), we estimate that the temperate and snowball planets have MLDs of 25 and 8 meters, respectively; i.e., the temperate planet has approximately 3$\times$ the thermal inertia of the snowball planet. 

The greater thermal inertia of the temperate planet is primarily due to the presence of surface liquid water.  Following \cite{Pierrehumbert_book}, Earth's oceans have an average mixed layer depth of 20--50~m, solid surfaces (continent or sea ice) have effective MLDs of 2--3~m, and the atmosphere contributes another 2--3~m. Accounting for the 57\% coverage of surface liquid water in our temperate model, one expects it to have 2.4--5.5 times greater thermal inertia than the snowball planet, in good agreement with our simulations.

\subsection{Obliquity Seasons}\label{obliquity_section}
\begin{figure*}[htb]
\begin{center}$
\begin{array}{cc}
\includegraphics[width=84mm]{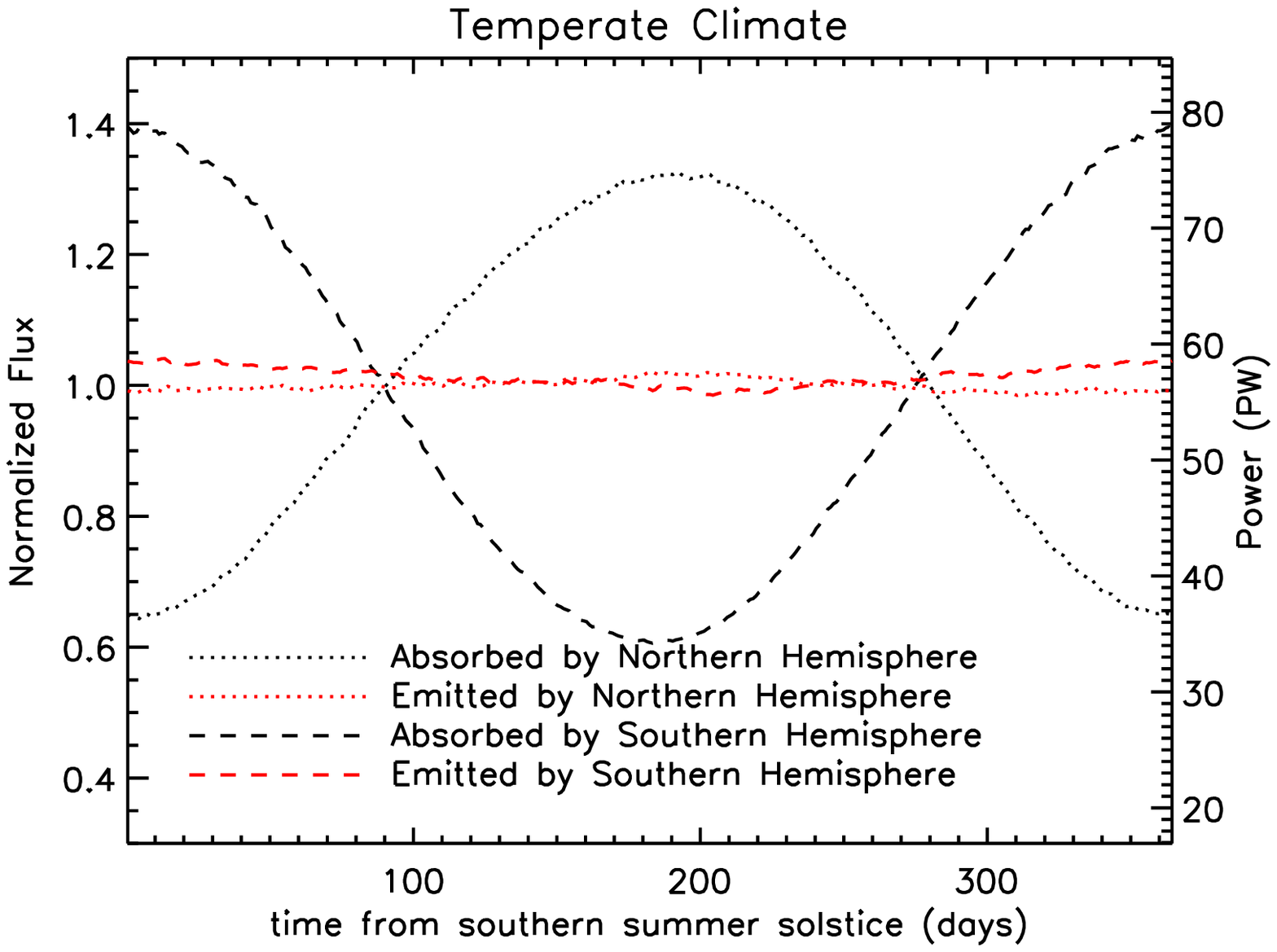} & \includegraphics[width=84mm]{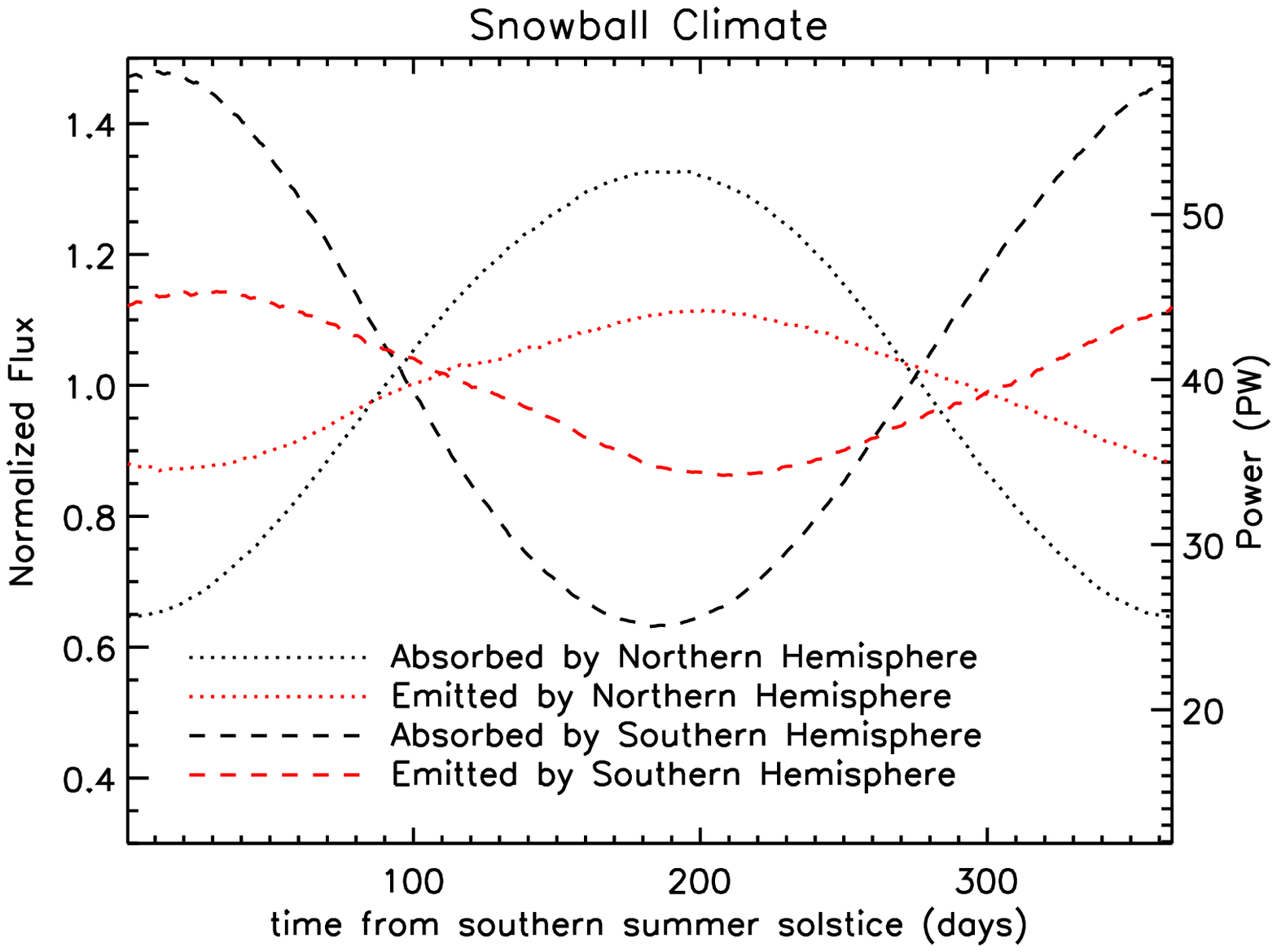}
\end{array}$ 
\end{center}
\caption{Absorbed (black) and emitted (red) flux for the northern (dotted) and southern (dashed) hemispheres in the temperate (left) and snowball (right) models.}
\label{obliquity_seasons}
\end{figure*}

In Figure~\ref{obliquity_seasons} we again consider total absorbed and emitted flux, but separate the northern and southern hemispheres in order to isolate the effects of obliquity. This approach tends to underestimate the magnitude of the seasons, because most of the flux is emitted from low-latitude regions that only experience small seasonal changes in incident radiation.  In fact, the deep tropics receive the most insolation at equinoxes, rather than summer solstice. Nevertheless, the seasonal forcing and response are clearly visible. 

The obliquity seasons are approximately $6\times$ greater for the snowball than for the temperate climate (26\% vs. 4\%, as shown by the red lines in Figure~\ref{obliquity_seasons}). If we treat the northern and southern hemispheres independently, ie: ignoring cross-equatorial heat transport, we may again use Equation~\ref{tau} to convert the absorbed/emitted amplitudes into thermal relaxation times. We obtain values of $\tau=1044$ and 155~days for the temperate and snowball climates, respectively. 

Since the global and obliquity seasons have the same forcing frequency, one expects the thermal inertia damping for both sorts of seasons to be the same. The large inferred value of $\tau$ for the temperate simulation obliquity seasons is a testament to the importance of heat transport from the summer to the winter hemispheres. Furthermore, since the obliquity seasons are so much weaker in the temperate simulation, one might conclude that its cross-equatorial heat transport is much greater than that of the snowball planet, but this is not the case. 

\begin{figure}[htb]
\begin{center}
\includegraphics[width=84mm]{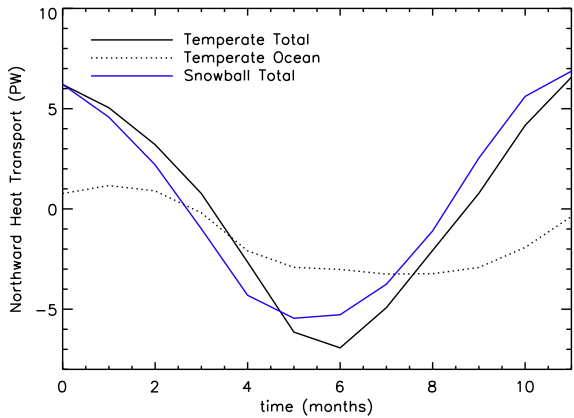} 
\end{center}
\caption{Seasonal variations in the cross-equatorial heat transport. The snowball planet has globally frozen oceans and thus negligible oceanic heat transport.}
\label{seasonal_transport}
\end{figure}

In Figure~\ref{seasonal_transport} we plot the seasonal cross-equatorial northward heat transport in both simulations. For the temperate planet, the ocean and atmosphere both play a role, while the frozen ocean in the snowball climate dictates that its heat transport is entirely atmospheric. Despite the lack of ocean transport, the seasonality of cross-equatorial energy transport is similar ---in absolute units--- for the two simulations. 

The temperate planet has a somewhat larger seasonal radiative forcing (40 PW vs.\ 30 PW; Figure~\ref{obliquity_seasons}) and much greater annual-mean heat transport than the snowball planet (cf. Figure~\ref{meridional_transport}).  In the absence of thermal inertia, one would therefore expect the temperate planet to transport much more energy from summer to winter hemisphere than the snowball.  But the greater thermal inertia of the temperate planet leads to a smaller cross-equatorial temperature gradient.  The greater heat transport efficiency and smaller temperature gradient conspire to produce nearly identical cross-equatorial energy transport in the temperate and snowball simulations.

The obliquity seasons have a much larger amplitude for the snowball planet than for the temperate planet, but the phase lag is nearly the same (1--2 weeks) in both cases.  This is partially due to the qualitatively different effects of heat storage vs.\ heat transport in damping obliquity seasons: heat storage tends to increase the phase lag while meridional heat transport does not \citep[this is manifestly true if heat transport is strictly diffusive, as in][]{Gaidos_2004}. 

The other reason that obliquity seasons can have very different amplitudes with similar phase lags is the inhomogeneous surface on the temperate planet. This is important for our temperate simulation because the equatorial landmass and polar ice conspire to leave only 57\% of the surface covered in liquid water. We find that the top-of-atmosphere thermal emission over sea ice exhibits large seasonal temperature variations that are approximately in phase with the forcing, while that over ocean exhibits muted seasonal variations.   As noted by \cite{Stine_2009}, mixing these signals results in a low amplitude, small-offset response, because the large amplitude component (solid surfaces) dominate the offset.  Since the land in our simulation is mostly tropical, its TOA thermal emission peaks during the relatively dry winter months, while the high surface temperature is masked by overlying clouds and humidity in the summer.   

\subsection{Diurnal Cycle}
\begin{figure*}[htb]
\begin{center}$
\begin{array}{cc}
\includegraphics[width=84mm]{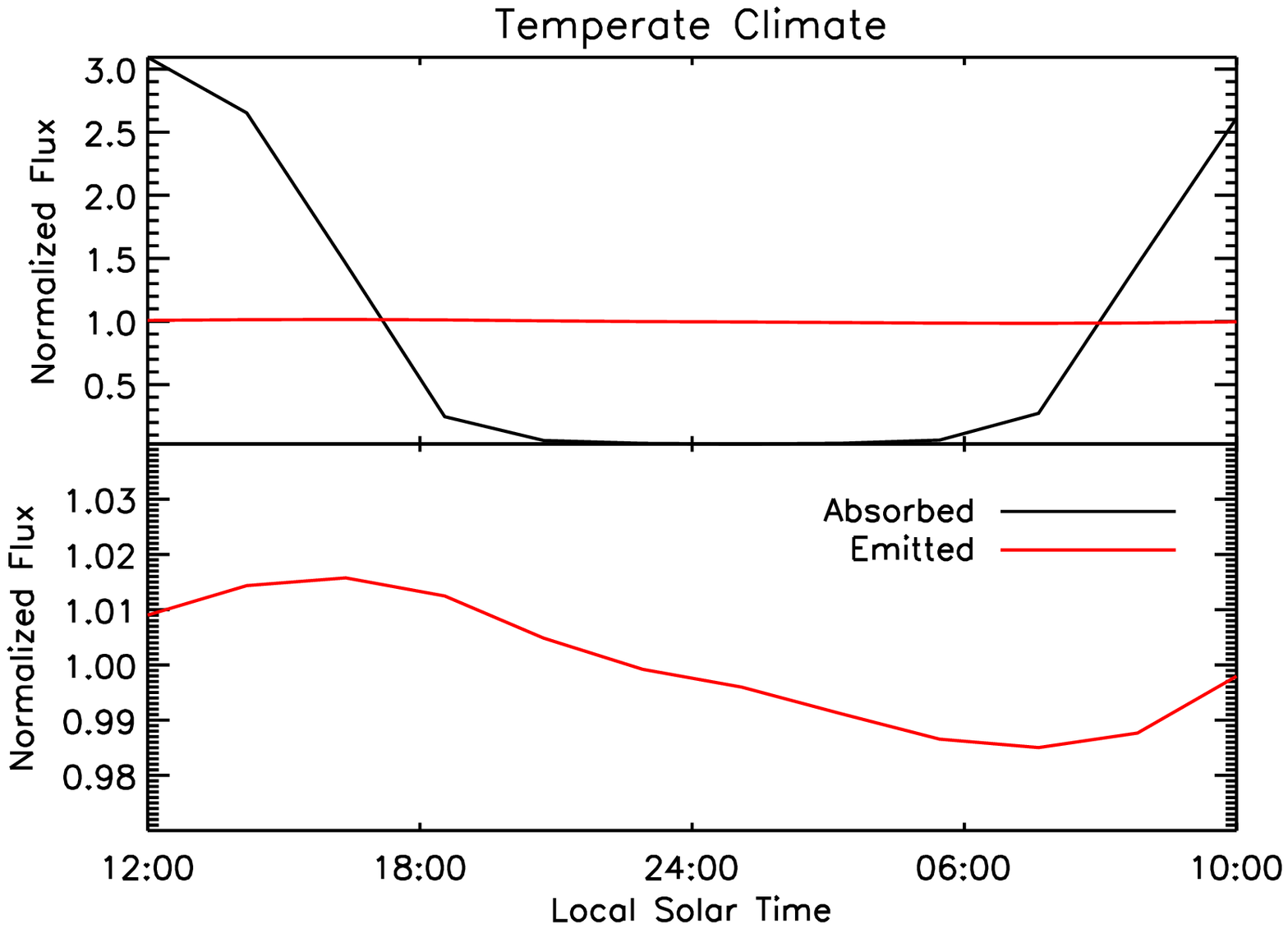} & \includegraphics[width=84mm]{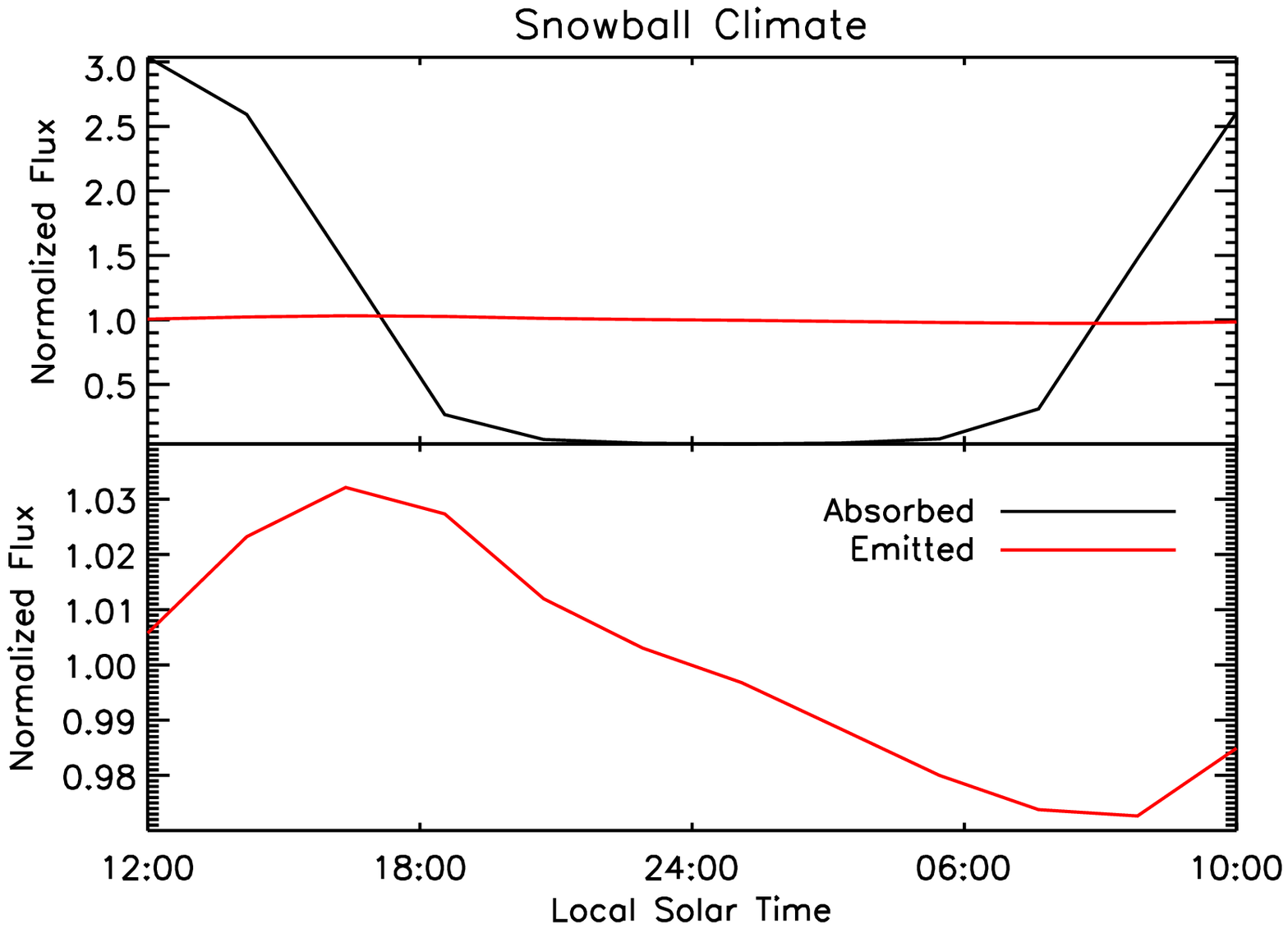}
\end{array}$ 
\end{center}
\caption{Absorbed (black) and emitted (red) flux as a function of local solar time.  The outgoing longwave radiation peaks in the late afternoon, while the lowest emitting temperature is reached shortly after sunrise. The angular curves are an artifact of the GCM's 2~hr output cadence.}
\label{diurnal_maps}
\end{figure*}

To isolate the diurnal cycle, we average over the entire planet and entire orbit, considering absorbed and emitted flux as a function of local solar time, shown in Figure~\ref{diurnal_maps}. For an Earth-like planet, zonal (E--W) heat transport is slow compared to planetary rotation, so the diurnal cycle is primarily damped by the planet's thermal inertia.

We again use Equation~\ref{tau} and obtain thermal relaxation times of $\tau = 16$ and 9~days for the temperate and snowball climates, respectively (or, equivalently, MLDs of 1.1 and 0.5 meters, respectively). These heat capacities are much smaller than those inferred from seasonal forcing because much less material senses the diurnal cycle. For land and sea ice, downward heat transport is conductive, so high frequency radiative forcing does not penetrate as deep as longer-period forcing, $\mu \propto 1/\sqrt{\omega}$ \citep{Pierrehumbert_book}. Upward heat transport in the atmosphere and downward heat transport in liquid water are not, strictly speaking, diffusive, but nonetheless only a relatively thin layer of the atmosphere and a shallow layer of the ocean respond to, and damp, the diurnal cycle.

Furthermore, the poor vertical ice resolution in our GCM tends to exaggerate the diurnal cycle in surface temperatures \citep{Abbot_2010b}. This will most affect the snowball simulation, leading us to underestimate its thermal inertia. 

The diurnal radiative forcing peaks at noon. In the high-frequency forcing limit, we expect phase-lags of $\pi/4$ and $\pi/2$ for diffusive and mixed-layer heat storage, respectively \citep[e.g., Section 7.4.2 of][]{Pierrehumbert_book}. The actual phase lag is somewhere in between these values.

\section{Disk-Integrated Photometry}
We now calculate time-resolved, disk-integrated photometry seen by a distant observer, ``light curves''. To compute the thermal light curves from the planet we assume diffuse emission and neglect limb-darkening:
\begin{equation} \label{observed_flux}
F = \frac{R^2}{\pi}\oint F_{\uparrow}(\theta,\phi, t) V(\theta, \phi, t) d\Omega,
\end{equation}
where $R$ is the planetary radius, $F_{\uparrow}$ is the top-of-atmosphere outgoing longwave radiation, $\theta$ is latitude, $\phi$ is longitude, $V(\theta, \phi, t)$ is the visibility of a given region of the planet for a given observer ($V$ is unity at the sub-observer point, drops as the cosine of the angle from the sub-observer point, and is zero on the far side of the planet), and $\oint d\Omega$ is the surface integral over a sphere.

Viewing geometry only enters Equation~\ref{observed_flux} through $V$, which is a function of the sub-observer location on the planet. Neglecting precession, the sub-observer latitude, $\theta_{\rm obs}$, is constant, while the sub-observer longitude evolves linearly in time, $\phi_{\rm obs} = \phi_{\rm obs}(0) - \omega_{\rm rot}t$, where $\omega_{\rm rot}$ is the planet's rotational frequency in an inertial frame (i.e., the sidereal rather than solar frequency).

\begin{figure*}[htb]
\begin{center}$
\begin{array}{cc}
\includegraphics[width=84mm]{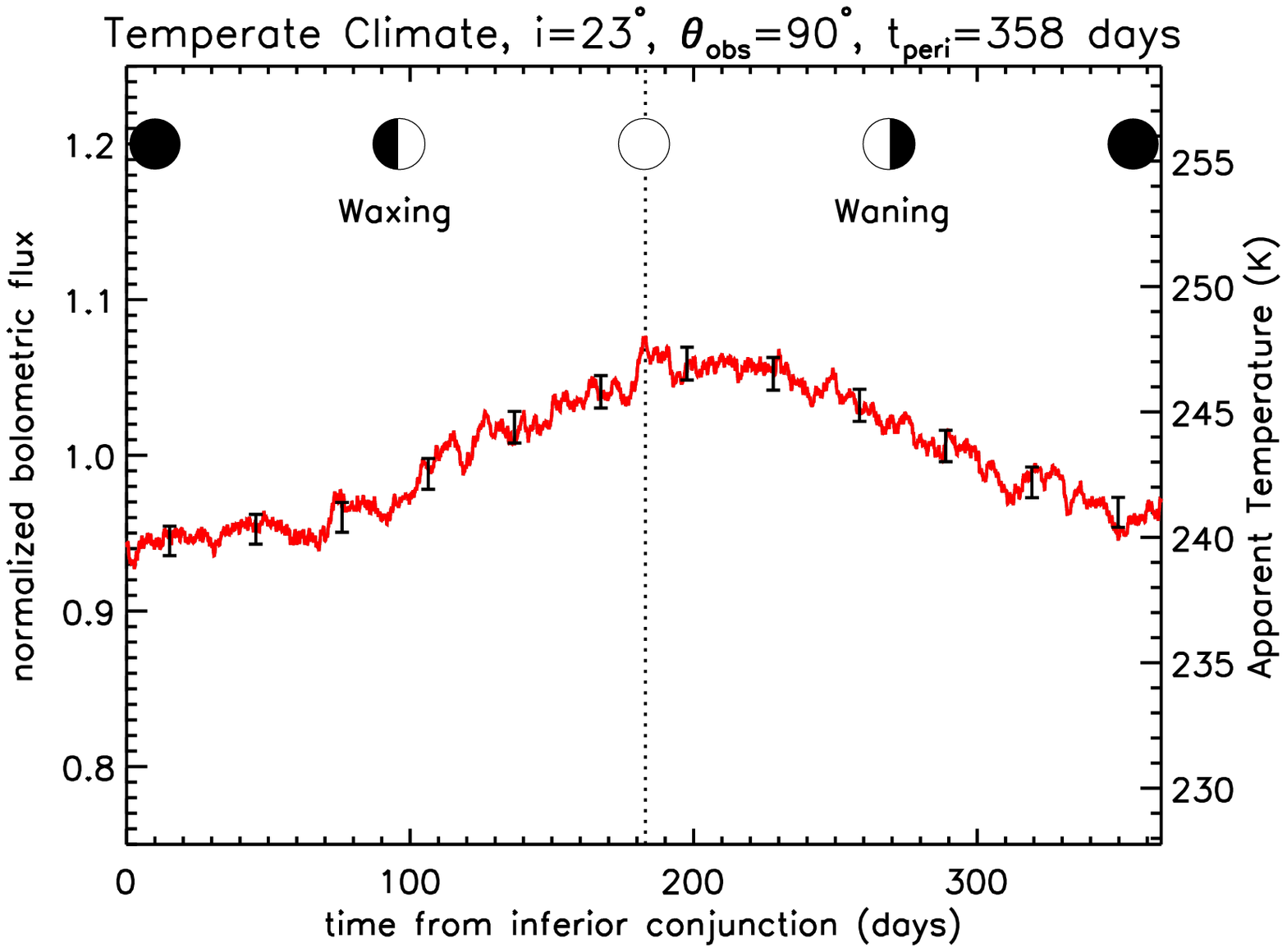} & \includegraphics[width=84mm]{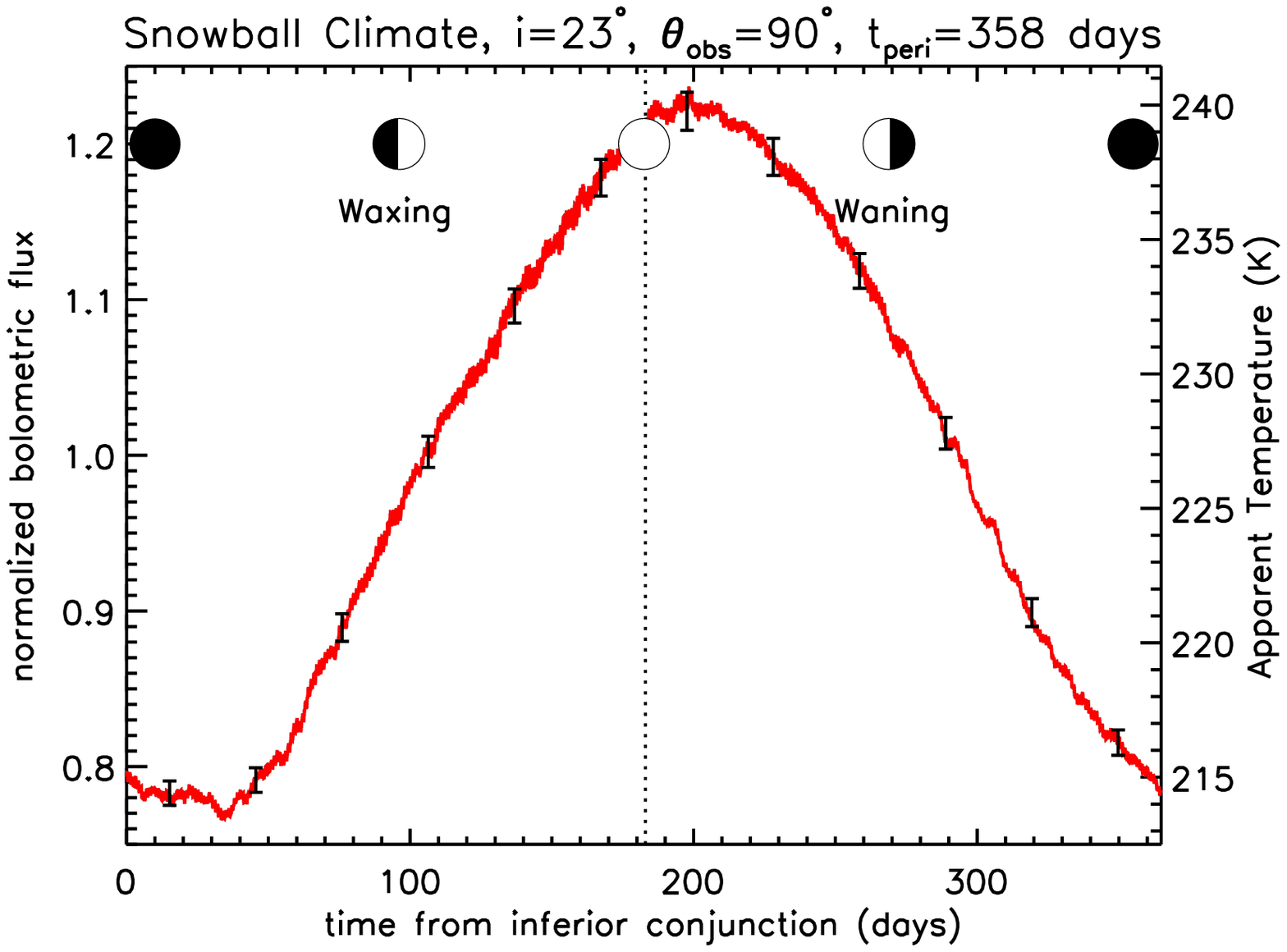}\\
\includegraphics[width=84mm]{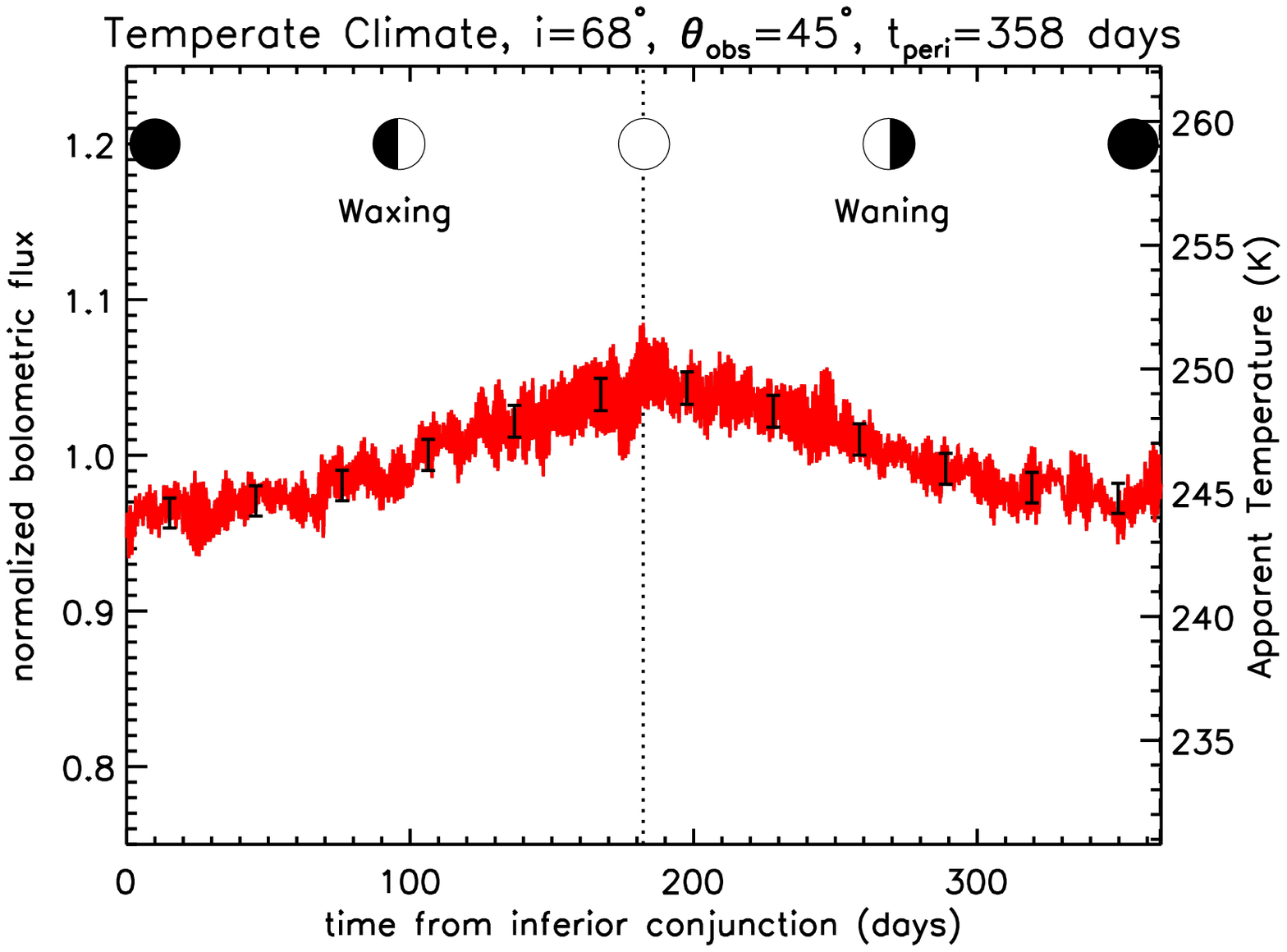} & \includegraphics[width=84mm]{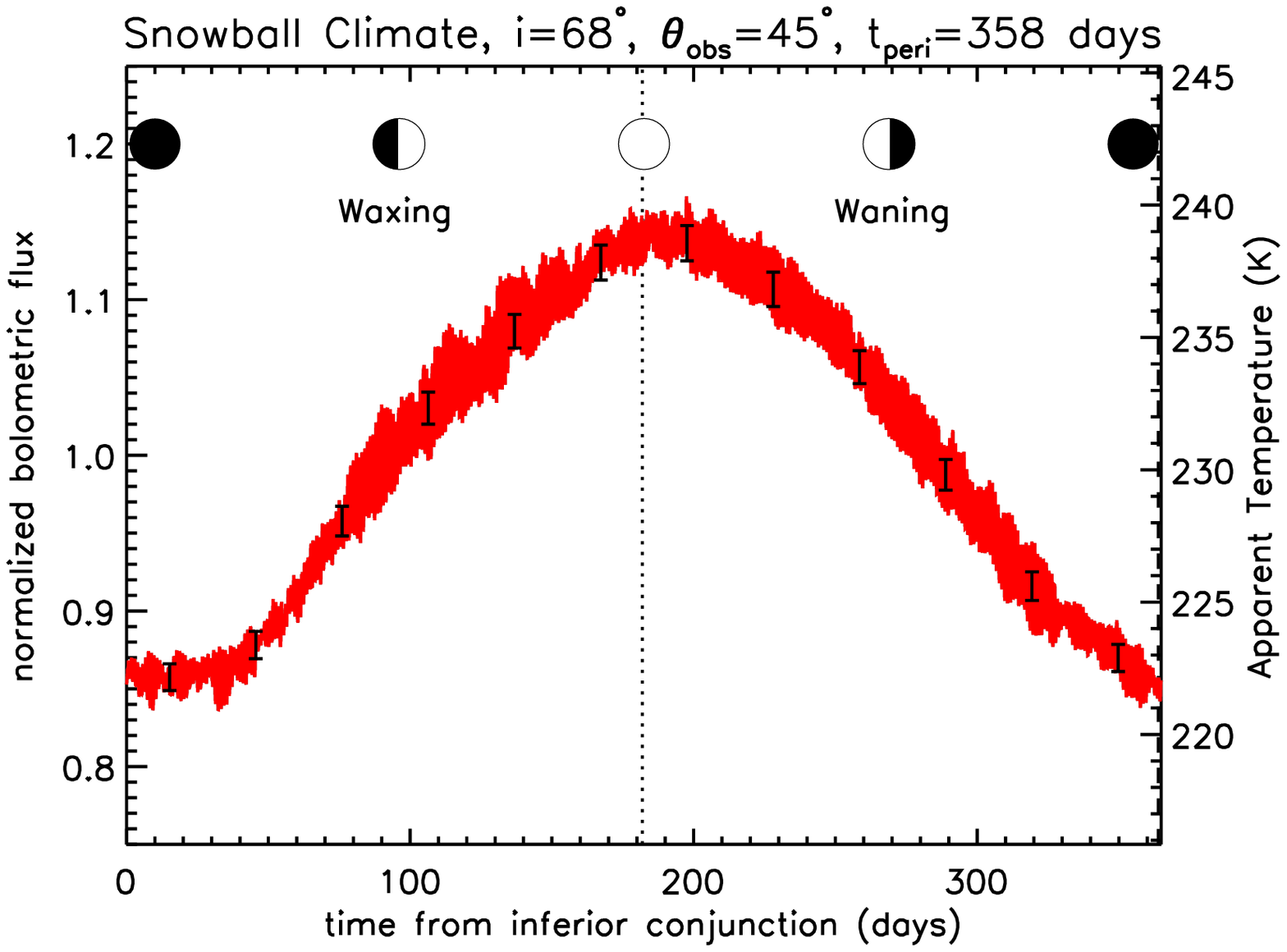}\\
\includegraphics[width=84mm]{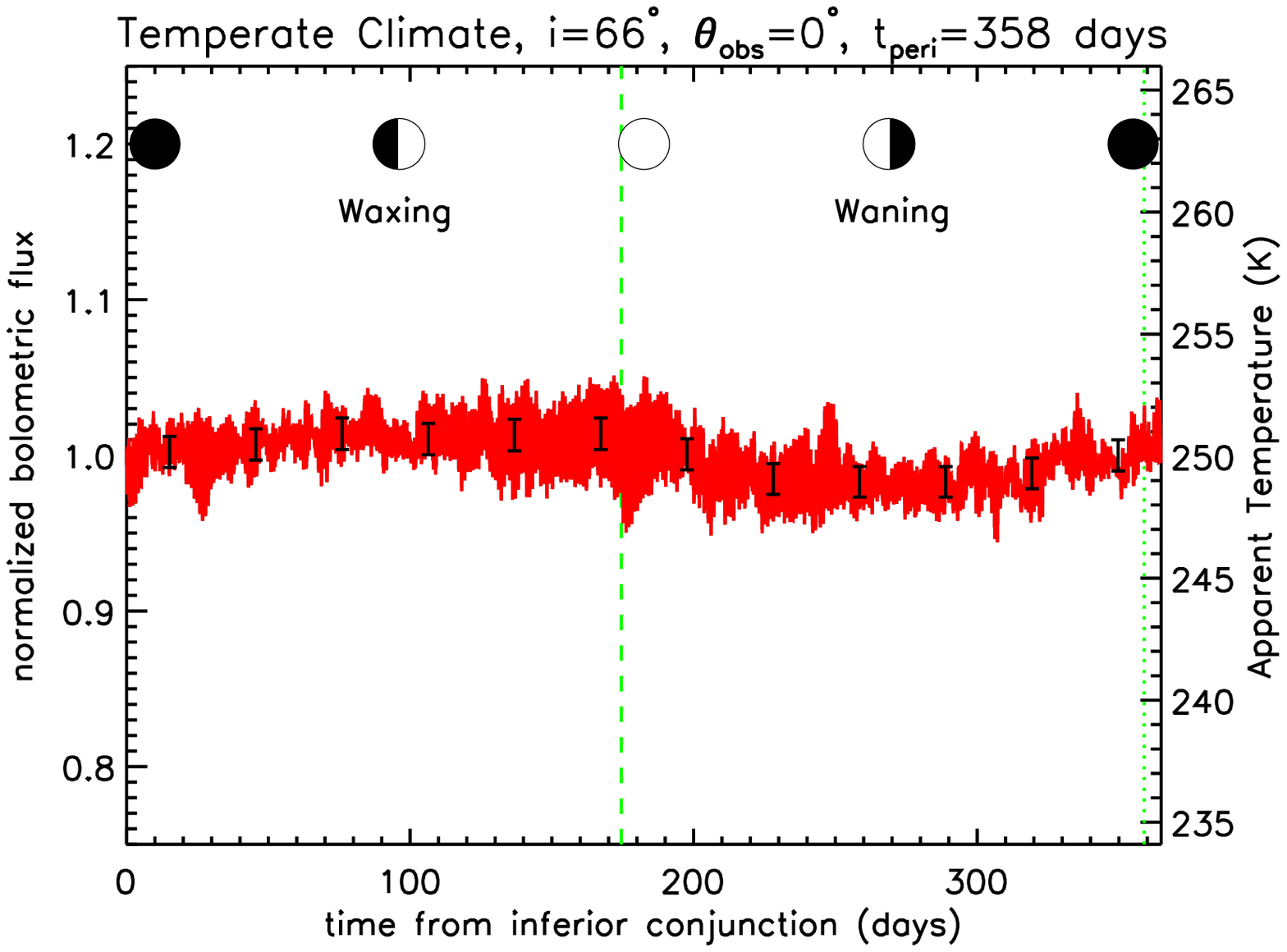} & \includegraphics[width=84mm]{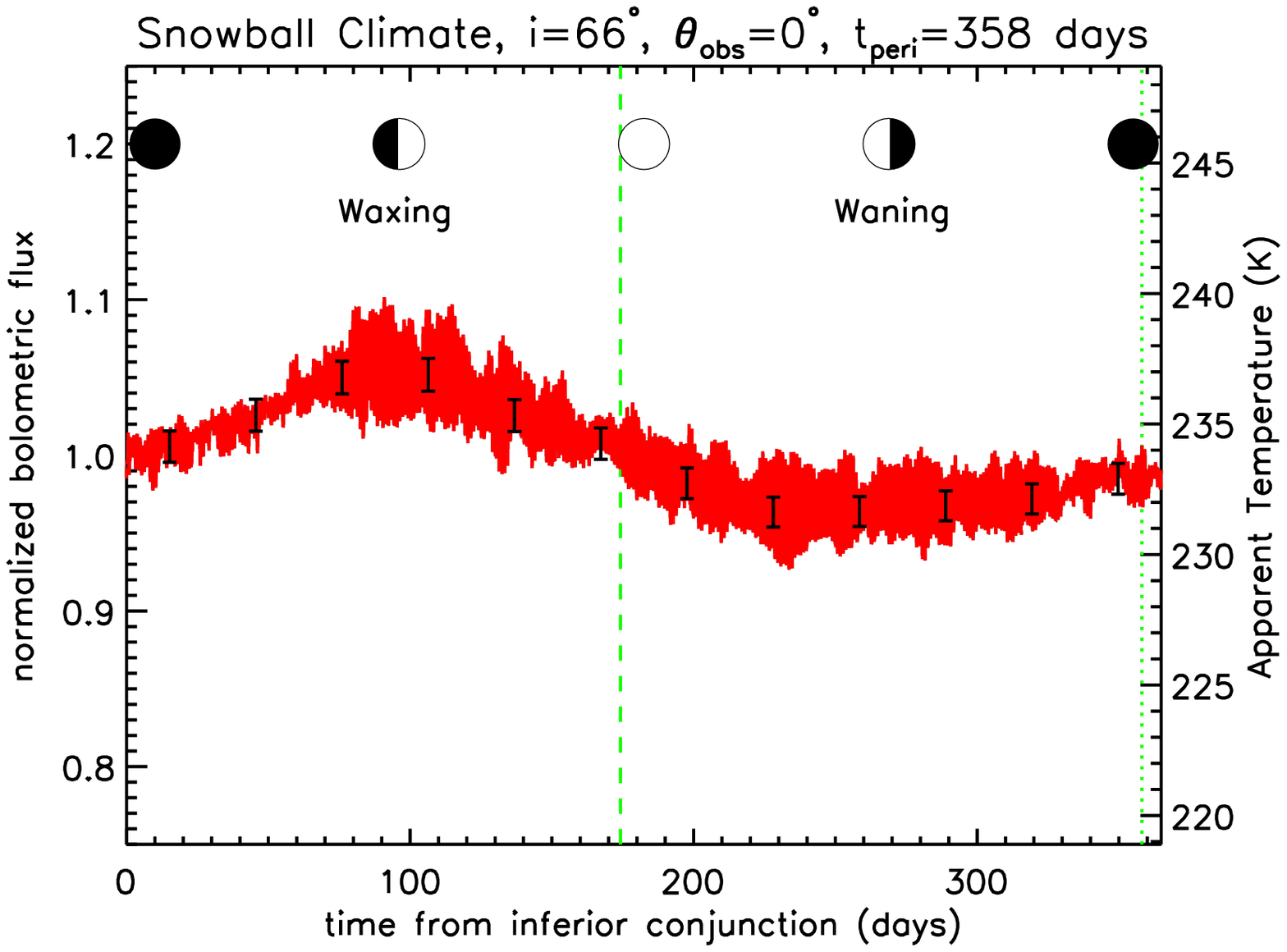}
\end{array}$ 
\end{center}
\caption{Thermal photometry as seen by a pole-on (top), mid-latitude (middle), or equatorial (bottom) observer. The black vertical dotted line shows northern summer solstice; it is plotted for the panels with thermal phases driven by obliquity seasons. The green vertical dotted and dashed lines are periastron and apastron, respectively; it is plotted in the panels where the thermal phase variations are most affected by eccentricity. We phase-fold 9 planetary orbits worth of simulated lightcurves; the inter-annual variability is $<0.1$\%. The black points show 1-month (rather than 2hr) integrations with illustrative error bars fixed at 1\%.  The cartoon along the top of each panel is illustrative: planets that are not in edge-on orbits do not exhibit new or full phases. The viewing geometry is described by the orbital inclination, $i$ (where $i=0$ is a face-on orbit and $i=90^{\circ}$ is edge-on), sub-observer latitude, $\theta_{\rm obs}$ ($90^\circ$ at the north pole, $0^\circ$ at the equator), and the time of periastron, $t_{\rm peri}$.}
\label{disk_averaged_lat}
\end{figure*}

For each viewing geometry we generate 10 years of simulated thermal phase variations and track the root-mean-squared orbit-to-orbit variability. This inter-annual variability is $\lesssim 1$\%, ranging between 0.2--1.5\% for most viewing geometries.  In Figure~\ref{disk_averaged_lat} we plot only one year of photometry.  For consistency, we always start the observations at inferior conjunction, when the planet is closest to the observer and we are seeing mostly its night-side.

The top, middle and bottom panels of Figure~\ref{disk_averaged_lat} show the same simulations, but seen by three different observers. 

The short-term variability in the thermal lightcurve is due to regions with different emitting temperatures rotating in and out of view.  These differences in temperature arise because of inhomogeneities in albedo, thermal inertia, and diurnal cloud cycles \citep[for a thorough discussion of diurnal cycles in Earth's OLR, see][]{Smith_2003}. Since we are considering integration times much longer than the planetary rotation rate, we do not study the rotational variability in this paper.  

Obliquity seasons are the largest-amplitude response in our simulations, so it is unsurprising that they are the dominant source of thermal phase variations for most viewing geometries.  The obliquity-driven phases are most extreme for a pole-on observer (top panels of Figure~\ref{disk_averaged_lat}) and most muted for an equatorial observer (bottom panels of Figure~\ref{disk_averaged_lat}). This is completely different from the inclination-dependance of diurnally-driven thermal phase variations, which are greatest for edge-on orbits and minimized for face-on orbits \citep[cf.][]{Cowan_2007, Crossfield_2010}.  

\subsection{High-Latitude or Mid-Latitude Observer}\label{hi_lat_sec}
Thermal phase variations are much greater for the snowball than the temperate simulation, as one would expect from Section~\ref{obliquity_section}. \cite{Gaidos_2004} showed that, in an idealized energy balance model, heat capacity is degenerate with obliquity in thermal phase variations. We cannot test this result because both of our simulated planets have the same obliquity, but the top four panels of Figure~\ref{disk_averaged_lat} shows that the actual situation is likely worse than they concluded.  The peak thermal emission occurs a couple of weeks after summer solstice for both climates, despite the muted amplitude of phase variations in the temperate simulation, because of meridional heat transport and inhomogeneous surface heat capacity (cf. Section~\ref{obliquity_section}). In other words, even if one knew the planetary obliquity and solstice phase independently, one would infer an incorrect thermal inertia from Equation~\ref{diff_equ}.

Notwithstanding the subtleties of phase lags, the amplitude of thermal phase variations roughly reflect the intrinsic strength of obliquity seasons. The amplitudes of intrinsic seasons differ by a factor of 6 for the two climates, while the obliquity-driven phase variations differ by 4.5. The midlatitude case ($\theta_{\rm obs} = 45^\circ$ in Figure~\ref{disk_averaged_lat}) is similar to the polar observer, but with somewhat muted phase variations.
 
\begin{figure}[htb]
\includegraphics[width=84mm]{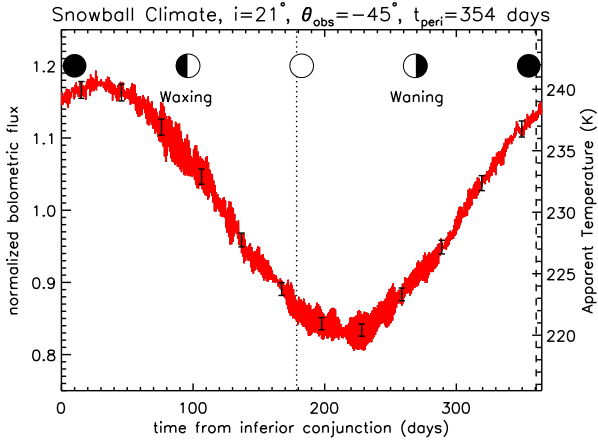}
\caption{Obliquity-driven phase thermal variations can peak near inferior conjunction, when we are mostly seeing the planet's night-side.  See Figure~\ref{disk_averaged_lat} for full caption.
}
\label{southern_phases}
\end{figure}

The phase variations seen by a high-latitude observer are dominated by obliquity seasons, but the peak flux roughly coincides with superior conjunction in Figure~\ref{disk_averaged_lat}. If one's intuition is based on thermal phase variations of hot Jupiters, it might seem that the maximum flux should always occur at/near superior conjunction, when we are seeing more of the planet's dayside. To dispel this notion, we plot the lightcurve for a southern observer in Figure~\ref{southern_phases}.  Inferior conjunction still occurs at $t=0$, but now it is the \emph{southern} seasons that affect the observed seasonal variability and the planet flux peaks near inferior conjunction. 

\subsection{Equatorial Observer}
For the equatorial view (bottom panels of Figure~\ref{disk_averaged_lat}), the planetary obliquity should not affect the thermal phase variations, since 1) the northern and southern hemispheres are equally visible, and 2) the continents are nearly equatorial, so there is little to distinguish northern summer from southern summer.  The phase variations from this vantage are therefore driven by global seasons and diurnal heating.  

If the albedo of the northern and southern hemispheres were very different, obliquity seasons would affect the phase variations seen by an equatorial observer. For example, while modern Earth's large N--S asymmetry in surface albedo is mostly obscured by the atmosphere and clouds \citep{Kasting_1993, Selsis_2007, Donohoe_2011, Voigt_2012}, the obliquity-driven global seasons are more important than the diurnal cycle for an equatorial observer \citep{Gomez-Leal_2012}. 

\begin{figure}[htb]
\begin{center}$
\begin{array}{c}
\includegraphics[width=84mm]{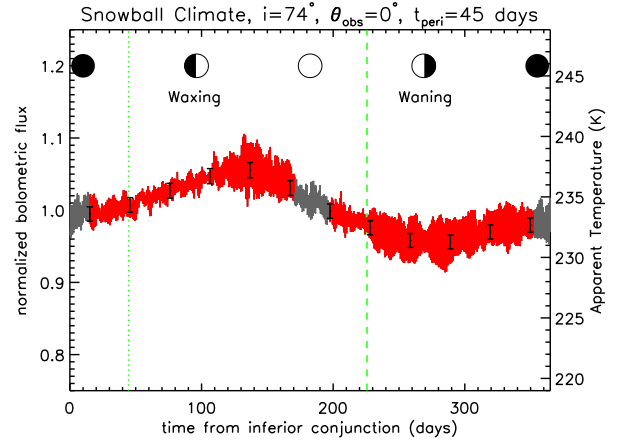}\\
\includegraphics[width=84mm]{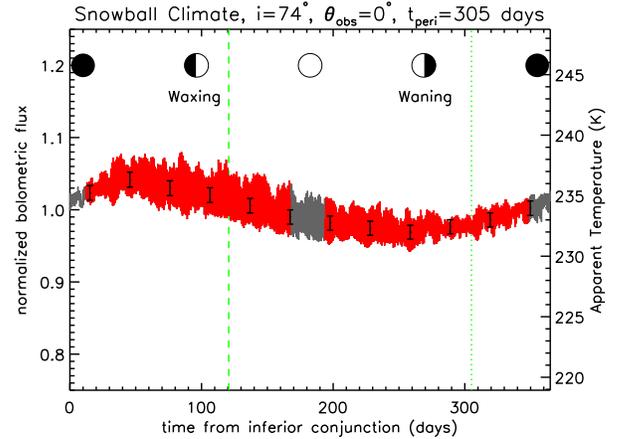}\\
\includegraphics[width=84mm]{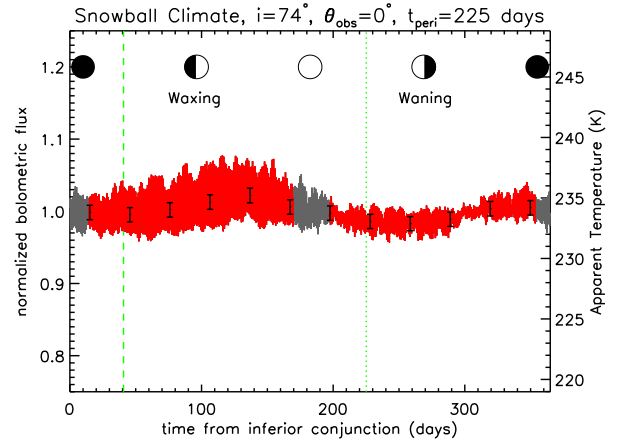}
\end{array}$ 
\end{center}
\caption{Thermal phase variations seen by an equatorial observer tend to be dominated by eccentricity and diurnal effects. Depending on the argument of periastron (shown here as the time of periastron, $t_{\rm peri}$), these two effects can interfere constructively (top panel) or destructively (bottom panel).  The grey regions denote the phases that will not be accessible because the planet will be inside the inner working angle (IWA) of the telescope; the geometry shown in the top panels is sufficiently close to face-on that the planet is never within the IWA.  See Figure~\ref{disk_averaged_lat} for full caption.}
\label{equatorial_phases}
\end{figure}

Figure~\ref{equatorial_phases} shows light curves for a snowball climate. We consider equatorial observers ($\theta_{\rm obs} = 0^\circ$) and fix the inclination at $i=74^\circ$, but vary the argument of periastron, which affects the amplitude and phase offset of thermal phases because of the competing effects of global seasons and diurnal heating \citep[cf.][]{Cowan_2011a}.

We gray-out the phases for which the planet would be inside the inner working angle (IWA) of a high-contrast imaging mission and therefore unobservable.  The IWA depends on the details of a mission's design and is a function of wavelength and distance. We adopt an IWA of 34~mas and an Earth analog at 10~pc, so that phase angles within 20$^\circ$ of full or new phase or inaccessible \citep[this is a reasonable IWA for TPF-I;][]{Lawson_2007}.  The planet is visible for the entire orbit if its orbital inclination is less than 70$^\circ$, as is the case for the geometries in Figures~\ref{disk_averaged_lat} and \ref{southern_phases}.

The changing power budget of the planet due to its non-circular orbit leads to a maximum in thermal emission near periastron (Figure~\ref{planet_averaged}). Diurnal heating, on the other hand, leads to a maximum in thermal emission between waxing quarter and superior conjunction, when afternoon local solar times of the planet are most visible (Figure~\ref{diurnal_maps}). Note that Earth has westward equatorial winds, but they are slow compared to the planet's rotation, so the hottest local solar time is still in the afternoon when averaged over the visible disk. For planets with thicker atmospheres, the wind direction becomes more relevant than the sense of the planet's rotation (e.g., super-rotation on Venus, Titan, or hot Jupiters). 

The thermal phase variations are therefore maximized when superior conjunction occurs a few months after periastron passage, so that the diurnal and eccentricity effects add up (top panel of Figure~\ref{equatorial_phases}); if the two effects are perfectly out of phase, the phase variations are attenuated (bottom panel). In all cases the phase variation amplitude is approximately twice as large in the snowball than the temperate simulation, as one would expect from Section~\ref{climate_forcing}.

The diurnal phase variations are smaller than the intrinsic diurnal cycle: even though a single local solar time is directly facing the observer at a given time, the disk-integrated lightcurve averages over many local times. Quantitatively, the first-order diurnal mode is attenuated by $\pi/4 \approx 0.79$ with respect to the time-constant component for an edge-on observer \citep{Cowan_2008}. Since eccentricity affects the entire planet's power budget, it does not suffer from this dilution. Therefore, while the intrinsic diurnal cycle is nearly twice the amplitude of the global seasons, the diurnal contribution to the phase variations is only $\sim 50$\% greater than that from eccentricity effects.

\section{Discussion}
\subsection{Feasibility}
Measuring thermal phase variations of an Earth analog will require space-based interferometry \citep[e.g., NASA's Terrestrial Planet Finder Interferometer or ESA's Darwin mission;][respectively]{Beichman_1998, Fridlund_2000}. \cite{Cockell_2009} estimated that \emph{Darwin} could obtain 20\% photometry on an Earth-sized planet at 10~pc in 12--24 hrs of integration. Extrapolating to month-long integrations, one expects 2.6--3.6\% photometry for an Earth-analog at 10~pc.  

We have simulated nine years of photometry, and found the inter-annual variability in disk-integrated flux is $\lesssim 1$\%, and a mere $\lesssim 0.1$\% after averaging over rotational variability. This opens the possibility of combining multiple orbits-worth of data.  For example, combining the nine years of photometry simulated here provides a three-fold enhancement in the signal-to-noise ratio, or 1\% photometry with 1-month integrations.

By using bolometric longwave radiation as a proxy for measured flux, we are assuming an instrument capable of capturing most of the planet's spectral energy distribution (SED).  This is a reasonable assumption because one will need to measure a directly imaged planet's SED in order to estimate its albedo and radius.  Considering bolometric flux also minimizes the impact of any unseen satellites, which only dominate the system flux in narrow wavebands \citep[][]{Robinson_2011}. Nevertheless, the presence of a large satellite would considerably complicate the interpretation of thermal phase variations.

When estimating the ability of TPF-Interferometer to establish a planet's orbit and residence in the HZ, \cite{Lawson_2007} assumed the planet has constant mid-infrared brightness throughout each orbit.  We find that the thermal flux from an Earth analog exhibits orbital variations of 5--50\%.  The amplitude of these variations is primarily governed by the planet's obliquity and the sub-observer latitude, neither of which will be known \emph{a priori} for directly imaged planets. This will complicate the retrieval of their orbital parameters from direct imaging. Note that the situation is yet worse at optical wavelengths: reflected flux variations are of order 100\%, since the planet's night side reflects no light.

\subsection{The Utility of Thermal Phase Variations}
There is an interesting parallel between non-transiting hot Jupiters and directly-imaged planets (in what follows we assume a circular orbit, but the argument can be generalized to non-zero eccentricity).  In the hot Jupiter scenario, the obliquity is assumed to be zero because of tidal effects, but orbital inclination is unknown. The thermal phase variations are entirely driven by diurnal heating, and therefore put a joint constraint on the planet's size, orbital inclination, and day--night temperature contrast \citep[e.g.,][]{Cowan_2007, Crossfield_2010}.  For directly imaged planets, the observer latitude is unknown, but the orbital inclination can be measured from astrometry.  For all but edge-on orbits, thermal phase variations can be assumed to be obliquity-driven, and therefore place a joint constraint on observer latitude and summer--winter temperature contrast.       

Focusing now on directly-imaged planets, the observer latitude is a function of one known quantity, the inclination, and two unknowns: the obliquity and equinox phase (Equation~\ref{lat_obs}). The summer--winter temperature contrast is a function of obliquity, thermal inertia and cross-equatorial heat transport. Two measured quantities, thermal phase amplitude and offset, are clearly insufficient to break the four-way degeneracy.\footnote{The distinction between heat storage and transport may seem academic, but in general they are only indirectly related.  On modern Earth, for example, storage is primarily in the oceans, while transport is mostly accomplished by the atmosphere.}  

Even without breaking any degeneracies, thermal phase variations of an imaged HZ planet might still be useful.  Small thermal phase variations would improve the odds of habitability, indicating either very weak radiative forcing, high thermal inertia and/or vigorous meridional heat transport.   This test does not work for a planet in a face-on circular orbit, which could simply have a small obliquity. Furthermore, as stated in the Introduction, large thermal inertia is no guarantee of habitability (e.g., Venus).

If the planet's obliquity and equinox phase were known through other means (see Section~\ref{constraining_climate}), then one could unambiguously estimate the summer and winter temperatures of the planet. The seasonal range in temperatures would be a useful climate constraint in and of itself, but one would like to constrain the intrinsic quantities. For a uniform planet, one could break the two-way degeneracy between thermal inertia and heat transport, since they have different phase offset signatures.  But this logic does not hold for an inhomogeneous planet because the mixing of signals from high and low thermal inertia regions complicates the interpretation of the thermal phase amplitude and offset. In principle it is possible to constrain inhomogeneous surfaces from rotational thermal variability, but this requires higher cadence and therefore a larger mission.

Insofar as the rotation rate of a terrestrial planet is determined by the final few major collisions of planet growth, it is natural to expect that some planets will rotate more slowly than Earth. Furthermore, planets on circular orbits about low-mass stars will have their planetary rotation rate slowed by tidal forces. In either case, the result would be stronger diurnal cycles than those considered in this paper.

Finally, many directly-imaged planets may have zero obliquity and hence no obliquity seasons. We have so far focused on true Earth analogs, terrestrial planets orbiting in the HZ of Sun-like stars.  But it is worth noting that, due the compact inner-working angle of an interferometric direct-imaging mission, many of its target stars will be K- and M-dwarfs.  Terrestrial planets in the HZ of such stars are likely to have had their obliquity tidally eroded \citep{Heller_2011} and therefore exhibit thermal phase variations driven only by eccentricity seasons and the diurnal cycle.  Such lightcurves would be much easier to interpret and might allow for a relatively robust determination of planetary heat capacity.

\subsection{Constraining the Climate of HZ Exoplanets}\label{constraining_climate}
We stated in the introduction that a planet's thermal inertia is an attractive ``second parameter,'' after albedo. We now expand on this idea, describing how various direct imaging observations constrain different key climate parameters, and how thermal phase measurements fit in the big picture. 

\emph{0-Dimensional Model:} The simplest possible climate model balances absorbed stellar flux with thermal emission, yielding a planet's emitting temperature.  This requires knowing a planet's albedo, which could be extracted from time-averaged thermal observations, but not from reflected-light observations because of the albedo-radius degeneracy.  This is of course insufficient to constrain surface conditions and hence habitability. 

\emph{1-Dimensional Model:} To estimate surface temperature, one needs to know the pressure of the emitting layer and surface, as well as the atmospheric lapse rate, the rate at which temperature increases as one descends in the troposphere. A thermal mission will detect any greenhouse gases since by definition they absorb in the thermal IR, and could measure surface pressure via pressure-broadened absorption features \citep[][]{Meadows_2010}. An optical mission may be able to detect some greenhouse gases in the near-infrared.  Significantly, a thermal mission can directly probe the surface temperature in opacity windows, providing an estimate of lapse rate and a check on climate models.  Optical data will contain, at best, indirect information about the planet's temperature.  

\emph{Higher-Order Model:} To construct a more comprehensive climate model, many additional planetary characteristics must be known. An optical mission could constrain rotation rate \citep{Palle_2008, Oakley_2009} and obliquity \citep{Kawahara_2010, Kawahara_2011, Fujii_2012}.  Furthermore, the presence of oceans could be inferred via their colors \citep{Ford_2001, Cowan_2009, Cowan_2011c, Fujii_2010}, polarization \citep[][]{Zugger_2010, Zugger_2011} and/or specular reflection \citep[][but confounding effects include clouds, Robinson et al.\ 2010, and snow, Cowan et al. 2012]{Williams_2008}. These quantities would improve our understanding of planet formation and geophysics, and would make it possible to convert observed thermal phase variations into estimates of intrinsic seasonal cycles. For planets theorized to have zero obliquity, thermal phase variations could constrain planetary thermal inertia.  But meridional heat transport and surface inhomogeneities would impede efforts to estimate planetary thermal inertia on planets with non-zero obliquity.

An exception for which the planetary thermal inertia can be estimated without resorting to periodic climate forcing is a planet where the thermal inertia is dominated by its atmosphere.  Recall that the lapse rate is a function of surface gravity and specific heat capacity: $\Gamma = -g/c'_{p}$, where the \emph{effective} specific heat capacity of the planet, $c'_{p}$, includes the effects of latent heat.  The thermal inertia coefficient for a well-mixed atmosphere is $\mu = c'_{p}p_{s}/g$, where $p_{s}$ is the surface pressure, and the latent heat effects of water vapor are again included implicitly.  A thermal mission of sufficient spectral resolution to measure pressure broadening could therefore estimate an exoplanet's atmospheric thermal inertia, provided that the planetary mass, and hence surface gravity, were otherwise constrained. 

\section{Conclusions}
Each of the three periodic climate forcings ---diurnal cycle, as well as obliquity and global seasons--- contribute to the thermal phase variations of habitable zone planets.  The damping for each of these oscillations is different.  The result is that while time-averaged thermal observations will powerfully constrain the climate of directly imaged exoplanets, phase variation measurements will be very challenging to interpret, unless planets can be assured to have zero obliquity. 

In a few limited cases, the lack of thermal variability on orbital timescales would provide circumstantial evidence of clement surface conditions.  On the other hand, very large swings in apparent temperature might discount a planet's habitability.

Our major conclusions are: 

1) Meridional heat transport can significantly affect obliquity seasons, and varies in relative strength from one planet to the next. Surface inhomogeneities in thermal inertia, meanwhile, lead to a mixing of high amplitude, small phase-lag signals with low-amplitude, large phase-lag signals. These complications make it impossible to quantitatively estimate a planet's heat capacity from obliquity-driven thermal phase variations, even if its obliquity is known through other means. 

2) Although the rotation rate of an Earth analog is shorter than the expected integration times of foreseeable thermal missions, the diurnal cycle shows up in the thermal phase variations on orbital timescales, and is the dominant effect for tropical observers. For rapidly rotating planets, the diurnal cycle and global seasons are both damped by heat storage alone, but planetary thermal inertia varies as a function of forcing frequency, whether heat storage operates as a mixed-layer or diffusively.

For a realistic habitable zone planet with an inhomogeneous surface, non-zero obliquity, and non-zero eccentricity, thermal phase variations are insufficient for measuring planetary thermal inertia.  

\acknowledgements
NBC acknowledges an insightful conversation about heat transport on snowball planets with D.~Koll, and an email exchange about obliquity erosion with R.~Heller. AV acknowledges support from the German Research Foundation (DFG) 
program for the initiation and intensification of
international collaboration. This work was supported by the Max Planck 
Society for the Advancement of Science.  All simulations were performed 
at the German Climate Computing Center (DKRZ) in Hamburg, Germany.

\end{document}